\documentclass[12pt]{article}

\usepackage{scicite}

\usepackage{times}

\usepackage{hyperref}
\usepackage{amsmath}
\usepackage{amsfonts}
\usepackage{graphicx}
\usepackage{comment}
\usepackage{booktabs}
\usepackage{threeparttable}
\usepackage{multirow}
\usepackage{mdwtab}
\usepackage{float}
\usepackage{siunitx}
\usepackage[normalem]{ulem}
\newcommand\redout{\bgroup\markoverwith
{\textcolor{black}{\rule[0.5ex]{2pt}{1pt}}}\ULon}
\usepackage{xcolor}

\newcommand{\suzuki}{Suzuki–Miyaura }
\newcommand{\bh}{Buchwald–Hartwig }
\usepackage{caption}
\newenvironment{sciabstract}{%
\begin{quote} \bf}
{\end{quote}}

\title{Large Language Models to Accelerate Organic Chemistry Synthesis}

\author{Yu Zhang$^1$, Yang Han$^2$, Shuai Chen$^2$, Ruijie Yu$^1$, Xin Zhao$^1$, \\Xianbin Liu$^1$, Kaipeng Zeng$^1$, Mengdi Yu$^1$, Jidong Tian$^1$, \\Feng Zhu$^{2\ast}$, Xiaokang Yang$^{1\ast}$, Yaohui Jin$^{1\ast}$, Yanyan Xu$^{1\ast}$\\
\\
\normalsize{$^{1}$MoE Key Laboratory of Artificial Intelligence, AI Institute,}\\
\normalsize{Shanghai Jiao Tong University, Shanghai 200240, P. R. China}\\
\normalsize{$^{2}$Frontiers Science Center for Transformative Molecules (FSCTM),}\\
\normalsize{Shanghai Key Laboratory for Molecular Engineering of Chiral Drugs,}\\ 
\normalsize{School of Chemistry and Chemical Engineering, Zhangjiang Institute for}\\
\normalsize{Advanced Study, Shanghai Jiao Tong University, Shanghai 200240, P. R. China}
\\
\normalsize{$^\ast$To whom correspondence should be addressed.}\\
\normalsize{E-mails: \{fzchem, xkyang, jinyh, yanyanxu\}@sjtu.edu.cn.}
}

\date{}

\begin{document} 
\baselineskip24pt
\maketitle


\begin{sciabstract}
Chemical synthesis, as a foundational methodology in the creation of transformative molecules, exerts substantial influence across diverse sectors from life sciences to materials and energy. Current chemical synthesis practices emphasize laborious and costly trial-and-error workflows, underscoring the urgent needs for advanced AI assistants. Nowadays, large language models (LLMs), typified by GPT-4, have been introduced as an efficient tool to facilitate scientific research. Here, we present \textit{Chemma}, a fully fine-tuned LLM with 1.28 million pairs of Q\&A about reactions, as an assistant to accelerate organic chemistry synthesis. \textit{Chemma} surpasses the best-known results in multiple chemical tasks, e.g., single-step retrosynthesis and yield prediction, which highlights the potential of general AI for organic chemistry. Via predicting yields across the experimental reaction space, \textit{Chemma} significantly improves the reaction exploration capability of Bayesian optimization. More importantly, integrated in an active learning framework, \textit{Chemma} exhibits advanced potentials of autonomously experimental exploration and optimization in open reaction spaces. For an unreported Suzuki-Miyaura cross-coupling reaction of cyclic aminoboronates and aryl halides for the synthesis of $\alpha$-Aryl N-heterocycles, the human-AI collaboration successfully explored suitable ligand (PAd\textsubscript{3}) and solvent (1,4-dioxane) within only 15 runs, achieving an isolated yield of 67\%. These results reveal that, without quantum-chemical calculations, Chemma can comprehend and extract chemical insights from reaction data, in a manner akin to human experts. This work opens avenues for accelerating organic chemistry synthesis with adapted large language models.

\end{sciabstract}



Chemical synthesis stands as the cornerstone of the discipline of chemistry, enabling the creation, exploration, and understanding of new substances~\cite{mendoza2012scalable,elvira2013past,ball2015chemistry,newman2021univariate, mikulak2020computational,jablonka2024leveraging}. Chemical synthesis stands as the cornerstone of the discipline of chemistry, enabling the creation, exploration, and understanding of new substances~\cite{mendoza2012scalable,elvira2013past,ball2015chemistry,mikulak2020computational}. In numerous cutting-edge scientific domains, chemical synthesis also plays a crucial role. For instance, in drug discovery, synthesis of complex molecules enables the creation of diverse candidates of small molecules with therapeutic potential~\cite{shen2021automation,chenthamarakshan2023accelerating}; in the realm of renewable energy, the design and synthesis of catalysts are essential for enhancing the efficiency of processes like hydrogen production, fuel cells, and carbon capture~\cite{tao2021nanoparticle,merchant2023scaling}.
Despite significant advancements in chemical instrumentation over the past few decades, the process of chemical synthesis continues to be laborious and time-consuming due to the vast reaction space and complexity of molecular structures~\cite{burger2020mobile, angello2022closed, betinol2023data, rinehart2023machine}. This iterative process typically involves reviewing the literature, designing reaction steps, and conducting wet lab experiments, which necessitates repeated trial and error~\cite{granda2018controlling, burger2020mobile}. Even with the aid of literature retrieval tools, for novel molecules, none of them has presented a valuable synthetic path toward integration of computer-aided synthesis planning (CASP) and expert effort~\cite{mehr2020universal,rohrbach2022digitization}. In summary, to realize efficient and highly selective chemical synthesis, there is an urgent need to reform the research paradigm of organic chemistry synthesis~\cite{sanchez2018inverse,wang2023scientific}.

Nowadays, the emergency of generative pre-trained transformer-based large language models (LLMs), typified by GPT-4, has triggered keen interest in leveraging such techniques to tackle scientific challenges~\cite{toniato2021unassisted,openai2023gpt, lehr2024chatgpt, kang2024chatmof}. Those include---but are not limited to---literature mining~\cite{dagdelen2024structured}, data annotation~\cite{hou2024assessing}, experimental planning~\cite{zheng2023gpt,boiko2023autonomous,canty2024reproducibility, ruan2024automatic}, scientific tools scheduling~\cite{zheng2023chatgptCenSci, m2024augmenting}, etc. Particularly in chemistry, Zheng et al. leveraged prompt engineering to instruct ChatGPT in the automation of text mining~\cite{zheng2023chatgpt}, experimental design~\cite{zheng2023gpt}, and multi-agents collaboration~\cite{zheng2023chatgptCenSci} for the discovery of metal-organic frameworks. Antunes et al. proposed CrystaLLM based on GPT-2 architecture to focus on modeling crystal structures through their textual descriptions~\cite{antunes2024crystal}. Jablonka et al. fine-tuned GPT-3 for inverse design and reaction classification tasks in organic chemistry~\cite{jablonka2024leveraging}. Further, other approaches have focused on designing chemical agents driven by LLMs to plan and conduct experiments~\cite{boiko2023autonomous, m2024augmenting, zheng2024integrating, ramos2025review}. Nevertheless, the chemical capabilities of GPT-4 remain limited to tasks such as molecular captioning, and translation, making it challenging to autonomously explore and optimize unreported reactions within an open reaction space.

In the field of organic synthesis, advanced AI techniques have made noteworthy progress. For seeking feasible routes, Coley et al. developed a preliminary platform that integrates retrosynthesis models and robotic apparatus~\cite{segler2018planning,coley2019robotic} to seek feasible routes; Shield et al. reported a framework of Bayesian optimization to identify high-yield experimental conditions within largely distinct parameter spaces~\cite{shields2021bayesian} for reaction condition exploration; Angello et al. reported a streamlined closed-loop workflow that leveraged machine learning algorithms to guide robotic experimentation for the discovery of general reaction conditions~\cite{angello2022closed}; Tang et al. leveraged statistical modeling techniques to model the mechanism of the oxidative addition process~\cite{tang2023interrogating}. Wang et al. applied reinforcement learning bandit optimization to identify generally applicable conditions by examining experimental feedback~\cite{wang2024identifying}. Nevertheless, these methods rely heavily on density functional theory (DFT) calculations to parameterize molecules, and manually feature selection by experts' knowledge, which leads to molecular screening and prediction continues to be a trial-and-error process. Besides, these methods need a HTE platform to collect experimental data, which shows less generalization capabilities new reactions~\cite{raghavan2023dataset, frey2023neural}. More importantly, existing methods for reaction exploration aim at finding suitable conditions within experts' predefined reaction condition pools, a.k.a. closed reaction space. However, potential conditions variables from open reaction space with higher performance might be ignored.


Towards these points, in view that molecules can be expressed in sequence, and the reaction is described as a natural language in literature, LLMs can be a potential solution owing to the following advances: (i) by learning molecular representations from SMILES, LLMs can capture the atomic arrangement of molecules and understand their chemical structures; (ii) pre-trained with extensive reaction data, LLMs have the capability to learn relationships among reactants, products, and conditions in reactions, thereby acquiring chemical knowledge akin to the learning process of chemists; (iii) the generative capability of LLMs enables the design of novel molecules, for example, in the ligand recommendation task, thereby facilitating the discovery of new chemical reactions. To this end, we present Chemma, a fully fine-tuned LLM based on LLaMA-2-7b, as a generative assistant to accelerate organic synthesis chemistry. Chemma is trained to answer questions of chemists, functionalized as human-AI interactions and experiment assistant. Chemma tackles several primary tasks with simple chemistry prompts, including forward reaction prediction, retrosynthesis, reaction performance prediction, condition generation, and reaction exploration\&optimization. These abilities have been assessed with open benchmark data and wet experiments. Particularly, for an unreported Suzuki-Miyaura cross-coupling reaction of cyclic aminoboronates and aryl halides for the synthesis of $\alpha$-Aryl N-heterocycles, with the assistance of Chemma, we explored suitable conditions only within 15 runs. The results confirm the capability of Chemma to explore open reaction space. 

\section*{Design of Chemma as a chemistry assistant}

We adopt the simplified molecular-input line-entry system (SMILES) to represent chemical molecules as strings. Such a chemical language is integrated with the natural language in Chemma, which can interact with chemists as an assistant. Fig.~1A presents the four primary tasks with simple chemistry prompts, including forward reaction prediction, retrosynthesis, condition generation, and reaction performance (yield and selectivity) prediction. Via tackling these tasks, Chemma can assist Chemists in two scenarios: (1) \textit{human-AI interaction}: Chemma receives task-specific queries constructed via instruction prompts, and generates potential solutions; (2) \textit{experiment assistant}: We propose an active learning framework driven by Chemma to accelerate the reaction exploration and optimization.

To develop Chemma, we extract reaction data from publicly available sources, including the Open Reaction Database (ORD)~\cite{kearnes2021open} and the USPTO-50k dataset~\cite{uspto-50k} (Supplementary Note~1). We then design a novel prompt generation system to create diverse and task-specific instructions with the assistance of GPT-4. Fig.~1B illustrates a template of instruction prompts to construct training datasets for various chemical tasks (Supplementary Note~2). Further, we integrate Chemma into an active learning framework to facilitate reaction exploration and optimization (Fig.~1C). In this framework, chemists initially propose potential conditions based on their prior knowledge. Chemma then functions as a chemistry assistant, iteratively suggesting the next reaction condition by incorporating feedback from the collected wet experimental results. After a round of ``suggestion-feedback loop'', we fine-tune Chemma to better adapt to this specific reaction. The detailed workflow of the Chemma is presented in the ``Workflow of Chemma'' section of the Methods.


Next, we outline the model architecture and training strategies of Chemma in Fig.~2. We implement two critical techniques in the development of Chemma (Fig.~2A). We initially fine-tune the base LLaMA-2-7b model with multi-task Q\&A datasets, including forward prediction, retrosynthesis, and condition generation, resulting in Chemma-SFT. Specifically for the condition generation, we further derive Chemma-RM to identify ``optimal'' conditions for reaction optimization. To accomplish this, we construct pair-wise ranking Q\&A datasets, where each question is associated with multiple answers reflecting varying levels of preference. Additionally, for other chemical tasks, including prediction of yield or selectivity, we design a two-stage training strategy. The left panel of Fig.~2B presents the details of training Chemma-SFT for generation tasks, namely stage-1. We extract the embeddings of reactions in the well-trained Chemma-SFT. In stage-2, we build regression networks (domain projector layers) to predict the reaction yields or selectivities. A comprehensive description of the model architecture, including detailed methodologies is provided in the ``Model architecture'' section of the Methods and Supplementary Note 2.2.

\section*{Performance of Chemma on open benchmark data}

We benchmark our model on various publicly available datasets to illustrate that Chemma can answer a wide range of organic synthesis questions via interactive conversations. The testing questions cover the aforementioned tasks, including forward prediction, single-step retrosynthesis, condition generation, yield prediction, and selectivity (including regioselectivity and enantioselectivity) prediction. Note that, to mitigate potential data leakage issues, we build specialized Chemma model for each chemical task independently. All reactions for evaluation are posted in Extended Data Fig.~2, and the results are presented in Fig.~3.

\textbf{Retrosynthesis.} 
Retrosynthesis planning in organic chemistry aims at designing a pathway to synthesize the target product starting from a set of purchasable molecules. Single-step retrosynthesis is the first key step, which provides the most possible reactants, often without consideration of reaction conditions. Via expressing molecules as SMILES, Chemma models single-step retrosynthesis as a sequence-to-sequence problem. For a fair comparison, we build Chemma from a plain LLaMA-2 model and all methods are trained and tested on the same USPTO-50k dataset~\cite{lowe2012extraction} (40k reactions for training). In the dataset, 40,000 samples are used for model training, with 5,000 samples for model validation and another 5,000 for testing. We select multiple reputable methods published in recent years as baselines for numerical comparison. These solutions can be categorized into two paradigms, graph-based methods~\cite{tu2022permutation, sacha2021molecule, seo2021gta, somnath2021graphretro} and transformer-based methods~\cite{wang2021retroprime, wan2022retroformer, chen2021deep, yao2024node}. We measure the top-\textit{k} accuracy of the predictions, defined as the proportion of test cases in which the correct answer appears among the top \textit{k}. We also test the performance of general-purpose LLMs for the retrosynthesis task with the same testing set, including GPT-4, GPT-3.5, GPT-4o, and LLaMA-2-13B (Supplementary Note~3 and Supplementary Table~1).

It is clear that, with the configuration of reaction class unknown and template-free, Chemma achieves 72.2\% top-1 accuracy, confidently outperforming the baseline methods (Fig.~3A). Specifically, Chemma outperforms the state-of-the-art transformer-based methods, NAG2G (55.1\%)~\cite{yao2024node}, by 17.1\% in terms of top-1 accuracy. All detailed results can be seen in Supplementary Table~2. Further, based on the remarkable capability of Chemma on single-step retrosynthesis, we generate the multi-step retrosynthesis routes for five drug molecules (Supplementary Note 3 and Extended Data Fig.~3). In these test cases, Chemma can mostly generate proper synthetic steps in the entire route, including reductive amination, aryl substitution, or nitro reduction reaction.

\textbf{Ligand recommendation.} 
We investigate the performance of ligand recommendation under identical reaction conditions. The objective is to recommend or generate the ``optimal'' ligands with the highest yield under identical reaction conditions in a predefined reaction space. We select the Pd-catalysed imidazole C--H arylation reaction extracted from ORD~\cite{coley2017prediction} for numerical evaluation, which consists of boronic acid derivative, aryl halide, ligand, base, and solvent substrates (Fig.~3B). Given a specific reaction, Chemma recommends a ligand under a predefined solvent-base combination of conditions. 

As shown in Fig.~3C, we randomly select eight cases for evaluation. The bases, solvents and ligands can be found in Extended Data Fig.~2C, which has been annotated by `B',`S', and `L'. The red distribution illustrates the yield distribution under each base-solvent-ligand combination of conditions, written as \textit{B-S}. For example, in the top panel of Fig.~3C, under the combination of CsOAc and DMAc, Chemma identifies the XPhos ligand. In comparison with other ligands, the yield distribution under XPhos is clearly on the right.
We also analyze the recommendation performance between diverse ligands for each base-solvent combination. We can observe that, for 15 of the 16 base-solvent combinations, the recommended ligand performs best in terms of the median value of reaction yields, suggesting that Chemma can recommend a high-yield ligand.


\textbf{Yield prediction.} 
Yield prediction in organic synthesis frequently presents a significant challenge, whereas product selectivity tends to be influenced by a limited number of elementary steps, numerous on-cycle and off-cycle events significantly affect reactivity~\cite{coley2017prediction,ahneman2018predicting}. Primary yield prediction methods calculate DFT descriptors for each molecule, and fit yields with machine learning models~\cite{ahneman2018predicting, li2023reaction}. In contrast, without DFT descriptors, Chemma predicts the yields with the latent representation of reactions extracted from Chemma-SFT (Fig.~2B).

Here, we report that Chemma can be used to predict the yields of reactions collected through the HTE, electronic laboratory notebooks (ELNs), or literature-derived data. For HTE datasets, we select Pd-catalysed \suzuki and imidazole C--H arylation reactions (Fig.~3B, Extended Data Fig.~2B). Pd-catalyzed C--H direct functionalization has earned increasing interest in pharmaceutical development for its ability to generate molecule complexity without the need for pre-functionalized starting material~\cite{szymanski2023autonomous, wang2024identifying}. Each reaction in the datasets is associated with conditions including aryl bromides, ligands, bases, and solvents, and the corresponding yield. For ELN datasets, we select the Buchwald–Hartwig reaction reported in~\cite{saebi2023use}. The distributions of reaction yields on these three datasets are given in Extended Data Fig.~4A-C. For the \suzuki and C--H arylation reactions, we allocate 30\% reactions of the full capacity whose substrates are not included in the training set as an out-of-sample test set; For the \bh [ELN] reaction, we randomly split 30\% of the entire dataset for testing.

The first four panels in Fig.~3D present the results on the \suzuki reaction. In the dataset, there are 28 combinations of boronics and aryl halides and four of them are selected as testing cases. The other two panels present the results on the imidazole C–H arylation and Buchwald–Hartwig [ELN] reactions, respectively. Color bands in Fig.~3D depict confidence intervals of the linear fitting between observations and predictions. Overall, Chemma provides acceptable performance on the three reactions. The main reason is that Chemma has been pre-trained on a large number of reaction data in stage 1 (Fig.~2B), and endowed with a certain degree of chemical knowledge. Furthermore, the test-set root mean square error (RMSE) for imidazole C–H arylation reaction is 6.59\% with a coefficient of determination $R^2$ value of 0.74, and RMSE for Buchwald–Hartwig [ELN] is 6.56\% with $R^2$ value of 0.79. They are slightly weaker than the \suzuki reaction, possibly due to the imbalance of yield distribution, the diversity of products, and the sparsity of reaction conditions. 

Also, we examine the performances of models via different training/testing split strategies (Supplementary Note 4). Chemma's consistent superiority over random forest (RF)~\cite{ahneman2018predicting} across varying data proportions, emphasizing its more effective use of limited datasets. Besides, we also evaluate the performance of Chemma for yield prediction of the Pd-catalyzed carbonylation reactions using literature-derived data~\cite{li2024challenges}. The $R^2$ value of Chemma achieves 0.74, clearly surpassing the traditional machine learning methods (Supplementary Note 4 and Extended Data Fig.~5).

\textbf{Regioselectivity and enantioselectivity prediction.} 
We apply Chemma on both regioselectivity and enantioselectivity prediction tasks. For regioselectivity, we adopt the dataset created by Li et al.~\cite{li2020predicting}, which consists of reactions about radical C--H functionalization over heterocycles. The dataset provides the reaction SMILES including different arene scaffolds, substitution patterns, and radicals, along with the free energy barriers ($\Delta G$) calculated by DFT annotations. We randomly select 30\% of the original dataset as the out-of-sample test set. 
For enantioselectivity, we focus on chiral phosphoric acid–catalyzed thiol addition to \textit{N}-acylimines from Zahrt et al.~\cite{zahrt2019prediction}. The dataset includes 25 reactants and 43 catalysts, making up 1,075 reactions. We randomly select 600 reactions for training and 475 reactions for tests. 

On the regioselectivity dataset, the $R^2$, RMSE and site accuracy of the predicted results are 0.93, 0.74 $kcal \cdot mol^{-1}$ and 78.74\%, respectively, which is close to the performance of RF model~\cite{li2020predicting}. A case study for regioselectivity can be seen in Supplementary Note 5 and Extended Data Fig.~6. On the enantioselectivity dataset, Chemma achieves an $R^2$ of 0.89 and a RMSE of 0.25 $kcal \cdot mol^{-1}$, compared to the $R^2$ of 0.915 and RMSE of 0.197 $kcal \cdot mol^{-1}$ reported by Li et al.~\cite{li2023reaction}. Notably, the only input to our model is the SMILES expression of the molecules, without the selection of physical organic features, or DFT descriptors, suggesting that Chemma has learned representation of molecules equipped with chemical knowledge. 

In addition, we further evaluate the performance of Chemma for enantiomeric excess prediction. We focus on the enantioselective addition of diethyl zinc to benzaldehyde in the presence of a racemic catalyst (RC) and an enantiopure chiral additive (CA). Chemma outperforms all baseline models, indicating superior accuracy in predictions (Supplementary Note 5.2).

\section*{Chemma-generated data benefits yield prediction and reaction optimization}

We first examine that \textit{``can Chemma-generated data improves the yield prediction of RF model driven by DFT descriptors?''}~\cite{ahneman2018predicting}. To this end, we randomly allocate a 10\% subset of datasets as the test set. For the other $90\%$ training data, we presume that only a fraction of data can be observed (e.g., 5\%), and utilize Chemma to complete their yields (e.g., 85\%). The seven scenarios are illustrated in Fig.~4A. We use the RF and DFT descriptors to build a yield prediction model~\cite{ahneman2018predicting} and select two HTE datasets for testing, the Pd-catalysed cross-coupling Suzuki–Miyaura reaction~\cite{perera2018platform} and the Pd-catalysed Buchwald-Hartwig reaction~\cite{ahneman2018predicting}. 

Figs.~4B-C compare the performance of yield prediction with varying fractions of real and generated data on \suzuki and \bh reactions, respectively. \textit{Chemma-enhanced RF} approach takes 90\% data for training, including both real and generated data; \textit{RF without enhancement} approach only utilizes the partially observed real data for training; the horizontal green line indicates the performance reported in ~\cite{ahneman2018predicting} using \textit{RF with 90\% real data}. It is clear that \textit{Chemma-enhanced RF} outperforms \textit{RF without enhancement} particularly when the fraction of real data is as low as a mere 5\%. Notably, \textit{Chemma-enhanced RF} achieves an $R^2$ value of 0.53 (0.72) on the Suzuki (\bh) reactions even when utilizing only 5\% real data, which is closely approaching 0.6 (0.8) using \textit{RF with 90\% real data}.

In a word, results substantiate that generated data presents an advantageous alternative for experimental observations, which is particularly important for few-shot learning scenarios. Furthermore, with reliable generated data, Chemma can enhance chemists' prior knowledge of the yield distribution within the reaction space. Such prior insights are of significant value in the optimization of reactions. In view of this, we also examine that \textit{``can Chemma-generated data improves the efficiency of reaction exploration with Bayesian optimization?''} Based on that we propose Chemma-BO, a modified Bayesian optimization (BO) framework, integrating the Chemma-generated data. Chemma-BO is designed based on the BO framework reported in Shields et al.'s work~\cite{shields2021bayesian}. The detailed workflow of Chemma-BO is represented in Fig.~4D and the Methods.

Figs.~4E-F present the cumulative maximum observed yields for optimizations using three distinct approaches, Chemma-BO, GPT-4, and BO, on both Suzuki-Miyaura and Buchwald–Hartwig reactions. The curves show the average performance of ten times repeated experiments. Five experimental runs are conducted in each batch. Additionally, we also compare the capability of GPT-4 as agents for reaction optimization as reported by Boiko et al.~\cite{boiko2023autonomous}. 

For the Suzuki reaction (Fig.~4E), compared to BO, Chemma-BO obtains more significant observed yields despite that they have similar initial conditions. Further, Chemma-BO achieves 98.5\% yield within the first 15 experiments (three batches), while both BO and GPT-4 require at least 50 experiments to achieve similar results. 
For the Buchwald-Hartwig reaction (Fig.~4F), Chemma-BO surprisingly achieves 98.7\% within the first $10$ experiments. Within the first 25 experiments, Chemma-BO achieves 99.8\% yield, while BO requires at least 50 experiments to achieve similar performance. It is worth noting that GPT-4 is unable to identify an acceptable reaction condition after 100 experiments. Ultimately, GPT-4's averaged maximum yield is limited to approximately 93\%, which is significantly lower than the maximum yields of Chemma-BO or BO.

The suboptimal performance of GPT-4 can be attributed to that, GPT-4 is best at general tasks that involving natural languages. However, reaction optimization is a complex task that depends not only on general knowledge but also on understanding the underlying chemical rules and principles~\cite{guo2023can, m2024augmenting}. From the experimental results, we note that the GPT-4's performance on \bh reaction is much weaker \suzuki reaction. The primary reason is that \bh reaction includes more uncertain variables such as reactants, conditions, and products than \suzuki reaction. Given this complexity, prior knowledge provided to GPT-4 may not be sufficient to understand this type of reaction. In contrast, Chemma-BO and BO can still work on \bh reaction due to the introduction of DFT descriptors. Moreover, we can see that Chemma has the potential to hasten the optimization process of the baseline BO approach by introducing generated yields. 

 \section*{Exploring and optimizing open reaction spaces with Chemma}

 
Exploring reaction space is a pivotal task in organic chemistry synthesis. It holds the key to discovering new reactions and optimizing the known condition space, leading to a deeper understanding of reaction mechanisms. Recent work focuses on reaction exploration within predefined conditions, a.k.a. closed reaction spaces~\cite{wang2024identifying, granda2018controlling, shields2021bayesian}. However, these methods heavily rely on the prior knowledge of experts, limiting the discovery of novel reactions. In recent years, LLMs have showcased their capabilities of hypothesis-generating, thereby having the potential to facilitate scientific discovery~\cite{wang2023scientific}. Thus, we investigate that \textit{``can Chemma explores reaction without predefined condition pools, a.k.a. open reaction space?''}

In response to this, we position Chemma as a chemistry assistant and integrate it into an active learning framework for reaction exploration and optimization in both closed and open spaces (Fig.~5A and Extended Data Fig.~7). To start the active learning loop, Chemma first generates the initial reaction conditions. Next, in the ``suggestion \& feedback loop'', chemists iteratively follow Chemma's suggestion and conduct wet experiments. After a round of experiments, if the target yield is not reached, chemists can fine-tune Chemma to start a new round of exploration. This allows Chemma to learn the chemical knowledge of the new reaction. The detailed workflow is presented in the Methods.

Figs.~5B-C illustrate the results of exploring ligands on imidazole C--H arylation and \bh reactions via the active learning framework (Extended Data Fig.~2). Our objective is to explore a suitable ligand that yields a high outcome under varying combinations of other conditions. Results reveal that an acceptable yield can be achieved in the first round of active learning (round 0). The exploration process of Pd-catalysed imidazole C--H arylation and \bh [HTE] reactions are presented in Extended Data Fig.~8, respectively. The detailed Q\&As between the chemist and Chemma are presented in Supplementary Fig.~3.

For the Pd-catalysed imidazole C--H arylation reaction (Fig.~5B), Chemma generates a ligand, XPhos, under the combination of CsOAc-BuCN using zero-shot prompts for the first experimental run with a yield of 76.63\%. In the second run, keeping CsOAc-BuCN unchanged, we instruct Chemma with ICL prompts to generate a ``higher-yield'' ligand. As a result, Chemma proposes another ligand, CgMe-PPh, which impressively produces a higher yield of 91.19\%. As Chemma generates consistent CgMe-PPh across multiple attempts by ICL prompts, we proceed to randomly select a new base-solvent combination for the next round of ligand exploration. In the fourth run, Chemma suggests CgMe-PPh under the combinations of KOPiv-\textit{p}-Xylene, which achieves the yield of 100\%.

For the Pd-catalysed Buchwald-Hartwig reaction (Fig.~5C), Chemma generates a ligand, AdBrettPhos, under the combination of P\textsubscript{2}-Et--2,1-Benzisoxazole using zero-shot prompts for the first experimental run. The achieved yield for this generated ligand XPhos is 4.95\%. In the second run, keeping P\textsubscript{2}-Et-2,1-Benzisoxazole unchanged, we instruct Chemma with ICL prompts to generate a ``higher-yield'' ligand. As a result, Chemma proposes another ligand, tBuXPhos, which produces a higher yield of 21.66\%. In the 12th run, Chemma suggests tBuXPhos under the combinations of MTBD-Isoxazole, which achieves the yield of 84.42\%. Note that Chemma may propose conditions outside the closed reaction space. In such cases, we ask Chemma to iteratively regenerate a new ligand until it falls into the predefined space. We list several unreported ligands generated by Chemma in Extended Data Fig.~9 and synthesize three ligands (L1, L3, and L6) to verify their effectiveness. We observe that the yields of L1 and L3 are 6\% and 16\%, respectively, while the reactivity of the L6 remains notably weak. That is, Chemma faces challenges in generating ligand molecules with high reactivity for unseen reactions in its training stage. The active learning pipeline could be a solution, in which Chemma can be fine-tuned in the loop to enhance its capability to understand the reactions.

Furthermore, we utilize the active learning framework to explore open condition space for an unreported reaction, synthesis of $\alpha$-aryl N-heterocycle, with the goal of maximizing the yield. $\alpha$-Aryl N-heterocycles are prevalent structural motifs encountered in various natural products, pharmaceuticals, agrochemicals, and chiral catalysts~\cite{taylor2014rings}. However, surprisingly, efficient methods to access $\alpha$-Aryl N-heterocycles through Suzuki-Miyaura cross-coupling of cyclic aminoboronates and aryl halides have not yet been reported. It underscores the considerable demand for the synthesis of $\alpha$-aryl N-heterocycles, which are becoming areas of active research in organic chemistry~\cite{ma2020general, shu2022modular}. 

In order to synthesize this product, we explore the efficient ligands under the combinations of identical base-solvents within an open reaction space. The prompts that we interact with Chemma are shown in Extended Data Fig.~10. With the assistance of Chemma, we prioritize ligands because they often generate reactive intermediates with reactants (piperidin-1-yl benzoate), which affect the reactivity of aryl and nitrogen heterocycles coupling. Next, we expect to optimize the solvent with Chemma. 
Specifically in Fig.~5D, (1-benzoylpiperidin-2-yl)boronic acid ester and 4-bromobiphenyl are selected as substrates, and the initial combinations of starting condition variables are selected by chemists' experience~\cite{sarkar2022general}, which include base (NaOH), solvents (\textit{p}-Xylene), ligands and additives (cuprous oxide, water). We also ask Chemma to suggest proper ligands for the given reactants and product. As presented in \textit{round 0} in Fig.~5D, L1 and L5 are proposed by Chemma, and the other seven ligands are selected by chemists. Unfortunately, all ligands L1-L9, exhibit poor performance in the reaction, with yields consistently below 15\%. 

After iterative experiments in \textit{round 0}, we fine-tune Chemma by constructing pair-wise Q\&A ranking datasets. Subsequently, in \textit{round 1}, we ask Chemma to suggest three more potential ligands through a zero-shot interaction. Note that, following the suggestion of chemists, we change the solvent from \textit{p}-Xylene to 1,4-Dioxane, because of the low yields in \textit{round 0}. Chemma recommends the following three potential ligands, BreetPhos (L10), PhS-Phos (L11), PAd\textsubscript{3} (L12). To our delight, we find that tri(1-adamantyl)phosphine (PAd\textsubscript{3}, L12) performs better in the cross-coupling reaction, with the yield of desired 67\%. Subsequently in \textit{round 2}, further interacting with Chemma, we conduct more wet experiments with varying solvents including DME, MeCN, and THF. However, we find that 1,4-dioxane adopted in the previous round performs best among all evaluated solvents. 

Having established the acceptable conditions for substrates of aryl and nitrogen heterocycles, we test its efficiency on four more reactants, as presented in Fig.~5E. Regarding the scope of aromatic electrophiles, a wide range of aryl electrophiles, bearing both electron-donating and electron-withdrawing substituents, underwent C\textsubscript{sp\textsuperscript{2}}--C\textsubscript{sp\textsuperscript{2}} coupling with (1-benzoylpiperidin-2-yl)boronic acid ester to afford final products in 45\%-67\% isolated yields. In summary, the human-Chemma collaboration successfully explores suitable ligands and solvents within only 15 runs. The results confirm that Chemma can be utilized to explore suitable conditions in open reaction spaces, assisting chemists in synthesizing unreported compounds.

\section*{Discussion}
In the traditional chemistry research paradigm, chemists always acquire synthesis experience through reading reactions in the literature. Inspired by this, we propose Chemma to learn the chemical reactions in the form of natural languages. We find that Chemma achieves start-of-art performance in multiple chemical tasks and can effectively instruct chemical synthesis experiments.
Nevertheless, it is crucial to ensure responsible development and use of LLM-based models. Thus, we discuss the limitations and unintended risks of Chemma and propose possible mitigation strategies.

\subsection*{Limitations}

Although Chemma has been trained with a substantial amount of chemical reaction data, it is still a formidable task to tackle reactions with very limited observations. In this context, Chemma might generate suboptimal outcomes. To cope with this, developers can focus on improving the quality and diversity of the open-source chemical data and introducing abundant knowledge from experts into the LLM.
For an unreported reaction, Chemma is unable to generate effective answers without chemists' feedback (round 0 in Fig.~5D). In this case, we iteratively ask Chemma to generate the next reaction condition based on the feedback of the last wet experiment and fine-tune Chemma. After rounds of human-AI collaboration, Chemma can better understand the specific reaction. 
Moreover, when tasked with specific chemical problems, Chemma may generate diverse responses, resulting in additional effort to select reasonable answers. To alleviate this issue, we propose standardized templates of prompts for each task, and well-defined protocols for data collection and analysis (Supplementary Fig.~3). We will also integrate other expert-designed tools to ensure the robustness and accuracy of its chemical analyses.

As a LLM, Chemma inevitably faces hallucination concerns. Hallucinations could be a consequence of LLMs extrapolating beyond their training data. Especially in the exploration of open reaction spaces, Chemma might propose non-viable synthesis routes, or recommend incompatible conditions. To tackle this, Chemma should work as a copilot in the chemical laboratory governed by chemists. Further, hallucinations could also be addressed by introducing more reliable knowledge, including chemists' experience or chemical mechanisms. While on the other hand, some Chemma-generated hypotheses might be valuable, e.g. designing new catalysts or new routes for complex compounds (Extended Data Fig.~9). This could be an important subject for further study. 


\subsection*{Unintended risks}

While Chemma has demonstrated significant potential as a generative chemistry assistant, several unintended risks must be considered and managed. 

One of the primary risks associated with Chemma is its potential misuse. Although Chemma is specifically trained to assist chemists, there remains a possibility that it could produce harmful compounds. For example, individuals with malicious intent might leverage Chemma's capabilities to synthesize toxins, illegal drugs, or other hazardous materials. Therefore, in our released service (\href{https://ai4chem.sjtu.edu.cn}{https://ai4chem.sjtu.edu.cn}), we design hard-coded templates of prompts and instruction guidelines to ensure safety. Furthermore, we will incorporate automated chemical property screening algorithms to identify and block attempts to generate harmful compounds. In addition, we strongly recommend that the Chemma-generated protocols should be rigorously reviewed by experienced chemists before conducting any wet experiments.

Intellectual property is another concern for the responsible development of generative AI models such as Chemma. To address this issue, it is imperative to establish clear guidelines and policies concerning the ownership of synthesized chemical structures or materials generated by the models. We can introduce a comprehensive database comprising literature and patent information for post-processing. On the other hand, collaborating with legal professionals and industry stakeholders will aid in tackling these complex issues and instituting effective measures to safeguard intellectual property.

We strongly encourage users to critically evaluate the information provided by Chemma and corroborate it with established scientific literature and expert opinions to mitigate the risk of relying on a potentially biased generation. By integrating these approaches, developers can endeavor to minimize the impact of any deficiencies in Chemma's chemistry knowledge, thereby enhancing the overall effectiveness and reliability of LLM-powered chemistry assistants. 

\subsection*{Outlook}

In this paper, we present a fine-tuned large language model, Chemma, for organic synthesis chemistry. Compared to general-purpose LLMs, like GPT-4, Chemma is designed as a generative chemistry assistant. Chemma is capable of human-AI interactions for primary tasks in chemistry, including retrosynthesis, reaction performance prediction, condition generation, and reaction exploration \& optimization. These abilities have been assessed with open benchmark data and wet experiments. Especially, Chemma is able to explore open reaction spaces through an active learning pipeline. For a previously unreported N-heterocycles Suzuki reaction, Chemma explored optimized conditions within 15 runs, achieving a yield at 67\%. These capabilities underscore Chemma's design intent as a tool to facilitate and enhance the practice of synthetic organic chemistry.

Moreover, the chemical explanation of Chemma is another aspect worthy of investigation. Through pre-training on numerous literature corpus, and fine-tuned on specific chemical domains, Chemma can extract task-specific information from reactions, forming numerical features known as representations or embeddings. Next, we will investigate the downstream applications of embeddings for reactions or molecules, by explaining the chemical semantics of embeddings. 

Overall, our work highlights the value of adapted large language models for scientific research, opening up opportunities to accelerate organic chemistry synthesis. Moreover, by integrating this new scientific tool with embodied AI, there would be a great opportunity to construct an autonomous chemical laboratory, accelerating the discovery of new compounds and new reactions.

\section*{Methods}

In this section, we start by introducing the workflow of Chemma, highlighting its significance in the field of organic synthesis. We then provide a detailed introduction to the model structure and training strategy. Finally, we elaborate on our reaction optimization methods, including BO and the active learning framework, which collectively enhance the exploration and improvement of reaction conditions.

\subsection*{Workflow of Chemma}
Here, we complement some details of the workflow of Chemma, as delineated in Fig.~1. All these reaction data are collected from the literature, patents, and high-throughput experiments (HTE). The details of the datasets for training can be found in Supplementary Note~1 and Extended Data Fig.~1. For the construction of prompts, we generate 2,000 question prompt templates for each task using GPT-4 to construct the supervised fine-tuning dataset. These templates are carefully designed to ensure diversity, consistency, and completeness, providing a robust foundation for training the model. Details of the dataset introduction and construction process are provided in Supplementary Note 2, with representative examples of prompts provided in Supplementary Fig.~1. For the application of human-AI interaction, Chemma receives task-specific queries constructed via instruction prompts such as ``\textit{please give me an optimized ligand of this reaction [Reactant1.Reactant2\textgreater\textgreater Product]}'', and generates potential solutions. For the application of experiment assistant, Chemma is equipped with the capability to optimize reaction performance through the active learning framework. After optimizing Chemma to be adapted to the new reaction, we can interact with it by asking questions such as ``\textit{we already obtain the yield of this reaction is 11.94\%, could you please give me some new ligands that potentially get higher yields?}''. The framework works not only on predefined reaction spaces by experts, a.k.a., closed space, but also on open reaction space where the conditions are not limited to experts' prior knowledge. 

\subsection*{Model architecture}

Here, we introduce the details of the model architecture, as shown in Fig.~2. Chemma-SFT is achieved by fully fine-tuning the foundational LLaMA-2 model using datasets introduced in Supplementary Note 2. Chemma-RM is expected to predict the most effective conditions by learning the performance differences between various reactions. Thus, pair-wise ranking Q\&A datasets contain questions and answers with varying levels of preference. These levels of preference are determined based on reaction performance, including yield and selectivity. For example, given the Buchwald-Hartwig reaction (Fig.~2A), we employ two ligands XPhos and tBuXPhos, and the annotation will be marked as ``XPhos $>$ tBuXPhos'', indicating that XPhos is better than tBuXPhos for reaction optimization. Furthermore, we leverage reinforcement learning training strategy from human feedback (RLHF) and proximal policy optimization (PPO) algorithm to train Chemma-RM. The objective of Chemma-SFT is to minimize the cross-entropy loss between the actual next tokens and the predicted next tokens. The loss function for Chemma-RM is also based on cross-entropy, where comparisons serve as labels—specifically, the difference in rewards reflects the log odds that one response will be preferred over another by chemical experiment results.

Chemma-SFT and Chemma-RM can handle the generation tasks. For the regression tasks (performance prediction), we develop a two-stages procedure. In stage-1, we extract the embedding of a given reaction from the last hidden state layer of Chemma-SFT. For stage-2, we adopt a five multilayer perceptron (MLP) feedforward neural network to take in the embeddings of reactions and predict their yields or selectivities. We utilize the mean squared error function as a regression loss to optimize the networks.

For Chemma's training strategy, we perform full fine-tuning of the LLaMA-2-7B model for four epochs over approximately 72 hours using 8 $\times$ NVIDIA A800 GPUs. For multi-GPU training, we employ the distributed data parallel (DDP) training strategy. The detailed architecture of Chemma and training strategies is shown in Supplementary Note 2.

\subsection*{Implementation of Chemma-BO}
For a given reaction search space, Bayesian Optimization (BO) begins by collecting initial reaction yield data through an HTE platform. This data is then used to train a probabilistic surrogate model. Once the surrogate model is trained, new experiments in the reaction space are sequentially selected by optimizing an acquisition function designed to maximize the expected utility of candidate experiments for subsequent evaluation. The proposed experiments are then conducted, and the surrogate model posterior is updated. This process continues iteratively until the reaction yield is maximized, the resources are depleted or the space is explored to the degree that finding improved conditions is improbable.

Using the predefined settings of BO, our objective is to enhance reaction performance by adapting algorithm components critical to maximizing yield in reaction optimization. Firstly, we utilize Chemma to generate the yields for all reactions with varying conditions. Then, we select the Top-5 conditions based on Chemma-generated yields for chemical experimental validation. Notably, we directly acquire the observed yields of certain conditions from HTE for validating the performance of BO. However, during new experimental explorations, wet lab experiments are necessary to collect the corresponding yield data. In the follow-up steps, we build a probabilistic surrogate model, e.g., Gaussian process (GP), to fit the gap between the previously observed and Chemma-generated yields with the DFT descriptors. With the trained GP, we then update the predicted yields in the entire reaction space by adding the output of GP to Chemma-generated yields. Next, we select the top-5 conditions to rerun experiments and repeat the above steps until acceptable yields are reached. In summary, the classic BO utilizes a surrogate model to fit the observed yields, while the bias of Chemma-generated yields is fitted in Chemma-BO.

\subsection*{Active learning framework of Chemma for reaction optimization}
We position Chemma as a chemistry assistant and integrate it into an active learning framework for reaction exploration and optimization (Extended Data Fig.~7). Given a new reaction, chemists utilize instruction prompts to interact with Chemma within two reaction condition scenarios: (i) reaction space with predefined or condition compounds, a.k.a. closed reaction space; (ii) selecting some initial conditions as the first try without limiting the reaction space, a.k.a. open reaction space. In Extended Data Fig.~7, we ask Chemma to generate initial conditions by zero-shot prompts demonstration as round 0. Subsequently, we utilize generated conditions from Chemma to conduct the wet experiment and obtain observed yields. if experiment queries are reached, chemists can fine-tune Chemma with collected experimental data, which is stored in the electronic laboratory notebook (ELN), and start a new round of exploration and optimization. a.k.a. rounds 1-N. In round 1-N, we ask Chemma to recommend new conditions and obtain the observed yield through the wet experiment. If chemists are unable to achieve target performance or gather sufficient data in this round, they can continually generate new conditions. Alternatively, if they obtain ample experimental data without reaching promising performance, they can leverage experimental records to fine-tune Chemma and start a new round. Once chemists obtain acceptable performance, the optimization process is finished. In addition, the detailed prompts are shown in Extended Data Fig.~10 and Supplementary Fig.~3.

\section*{Data and model availability}
The data used for model training and testing are from open benchmark datasets (ORD and USPTO). The source data for retrosynthesis prediction are available via Zenodo at
\href{https://doi.org/10.5281/zenodo.15166275q}{https://doi.org \\ /10.5281/zenodo.15166275}. The datasets for yield prediction, including HTE, ELN, and literature data, were collected from~\cite{ahneman2018predicting, perera2018platform, saebi2023use, li2024challenges}. Data for regioselectivity and enantioselectivity prediction tasks were from~\cite{li2020predicting, zahrt2019prediction}. Chemma is available for free usage at 
\url{https://ai4chem.sjtu.edu.cn/}.




\bibliography{scibib}
\bibliographystyle{Science}

\section*{Acknowledgments}
We thank Prof. Kuiling Ding for valuable discussion on the design of this work, and thank SJTU AI for Science platform for computing support. This work was supported by the Shanghai Municipal Science and Technology Major Project (2021SHZDZX0102), and the Fundamental Research Funds for the Central Universities.

\section*{Author contributions}
YX and YZ conceived the research and designed the analyses. YZ designed and implemented the Chemma model. YH, SC, FZ and YZ performed the wet experiments. YZ, YX, RY, XZ, XL and KZ processed the data and performed the results analyses. MY and JT discussed the model design. YJ and XY built the computing platform for model training. YZ, YX and FZ wrote the paper. YX, FZ, YJ and XY supervised the research.

\section*{Inclusion \& ethics}
All contributors who fulfill the authorship criteria are listed as co-authors in this paper. Other contributors who do not meet all criteria for authorship are listed in the Acknowledgements.

\section*{Competing interests}
The authors declare that they have no competing interests.


 



\newpage
\begin{figure*}[htb!]
\centering
\includegraphics[width=1.0\linewidth]{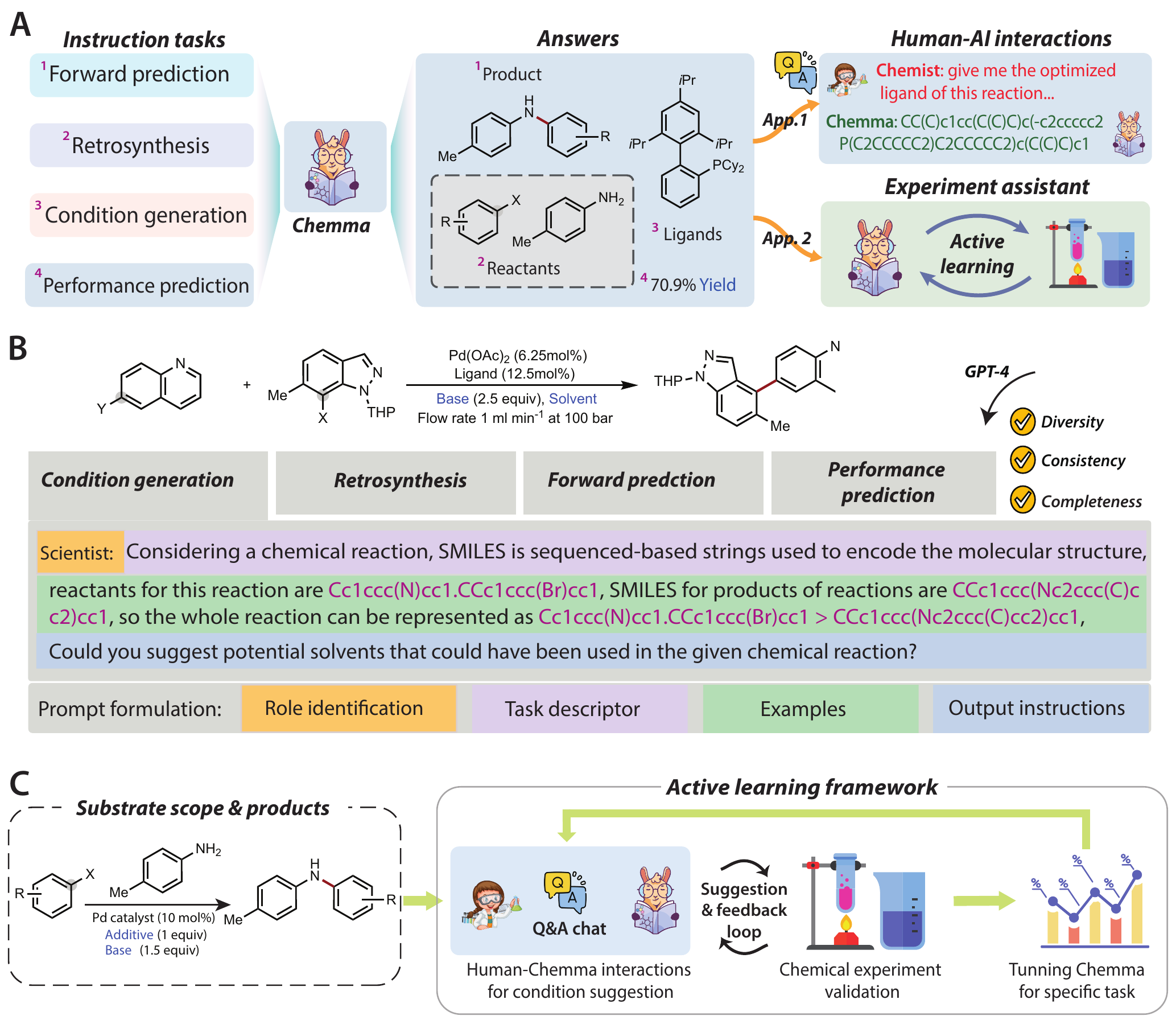}
\label{fig1:architecture}
\end{figure*}
\noindent{ \textbf{Fig.~1. Functions and applications of Chemma.} 
(\textbf{A}) Illustration of using Chemma to accelerate organic chemistry synthesis. Scientists can interact with Chemma about four main tasks, including forward prediction, retrosynthesis, condition generation, and performance prediction (e.g., yields and selectivities). We also present an active learning framework, using Chemma as an experiment assistant, to accelerate the exploration of new reaction spaces.
(\textbf{B}) Prompt templates for Q\&A pairs preparation. The construction of supervised fine-tuning datasets involves generating 2,000 question prompt templates for each task using GPT-4, ensuring the diversity, consistency, and completeness of the training datasets. 
(\textbf{C}) Active learning framework for experimental guiding. Chemists can interact with Chemma in the generally applicable close-loop workflow based on an active learning framework to explore new reaction space.}

\newpage
\begin{figure*}[tb!]
\centering
\includegraphics[width=1.0\linewidth]{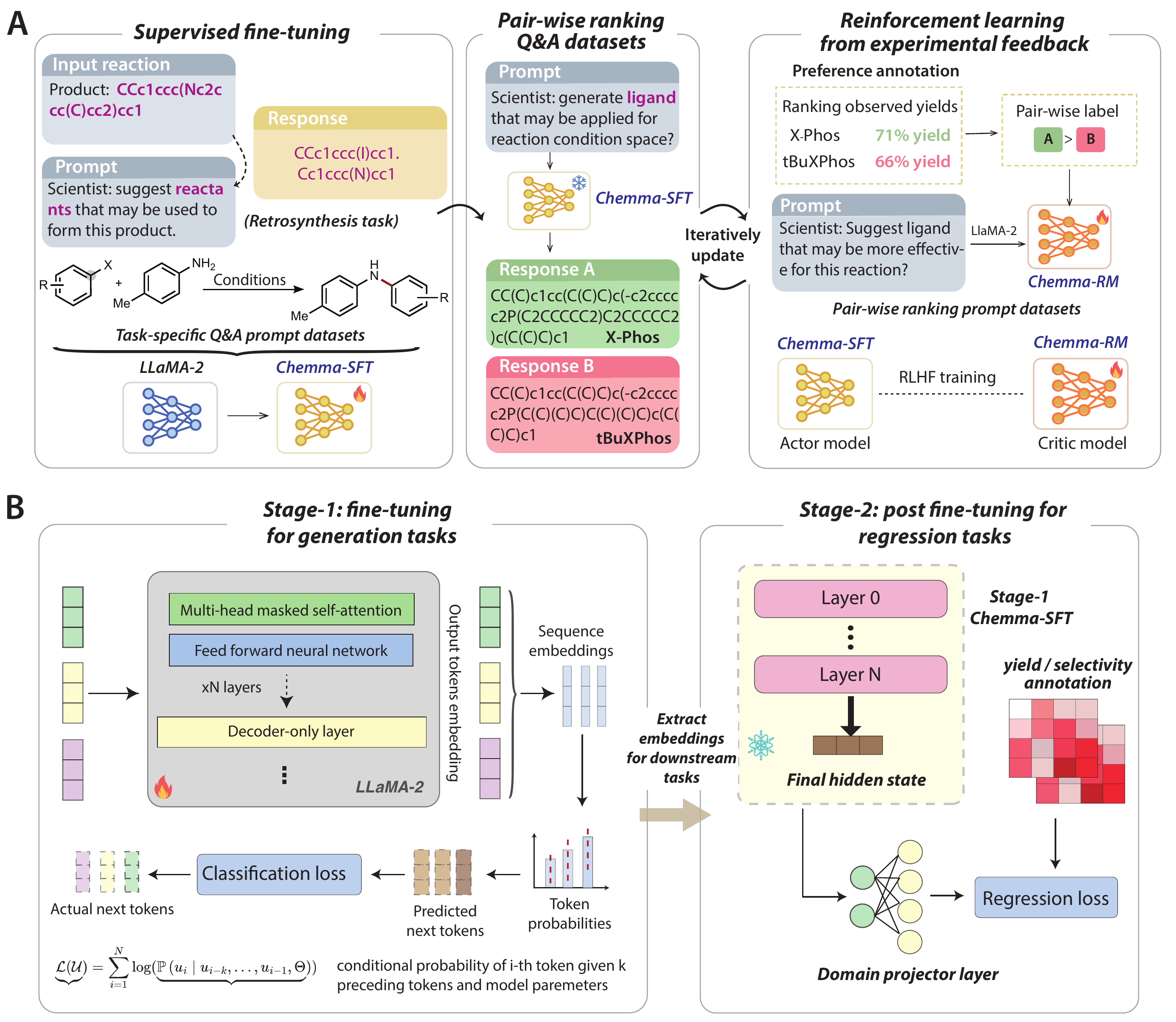}
\label{model-overview}
\end{figure*}
\noindent{{\bf Fig.~2. Model architecture of Chemma and training strategies for chemical tasks.} (\textbf{A}) Model architecture of Chemma. We first develop Chemma-SFT by fine-tuning the base LLaMA-2-7b model with multi-task Q\&A datasets, including forward prediction, retrosynthesis, and condition generation. Specifically for the condition generation, we further derive Chemma-RM for reaction optimization. (\textbf{B}) Two-stage training strategy for reaction performance prediction. In stage-1, we extract the embeddings of reactions in a well-trained Chemma-SFT. In stage-2, we employ MLP networks for regression tasks, including yield and selectivity prediction.}

\newpage
\begin{figure*}[htb!]
\centering
\includegraphics[width=1\linewidth]{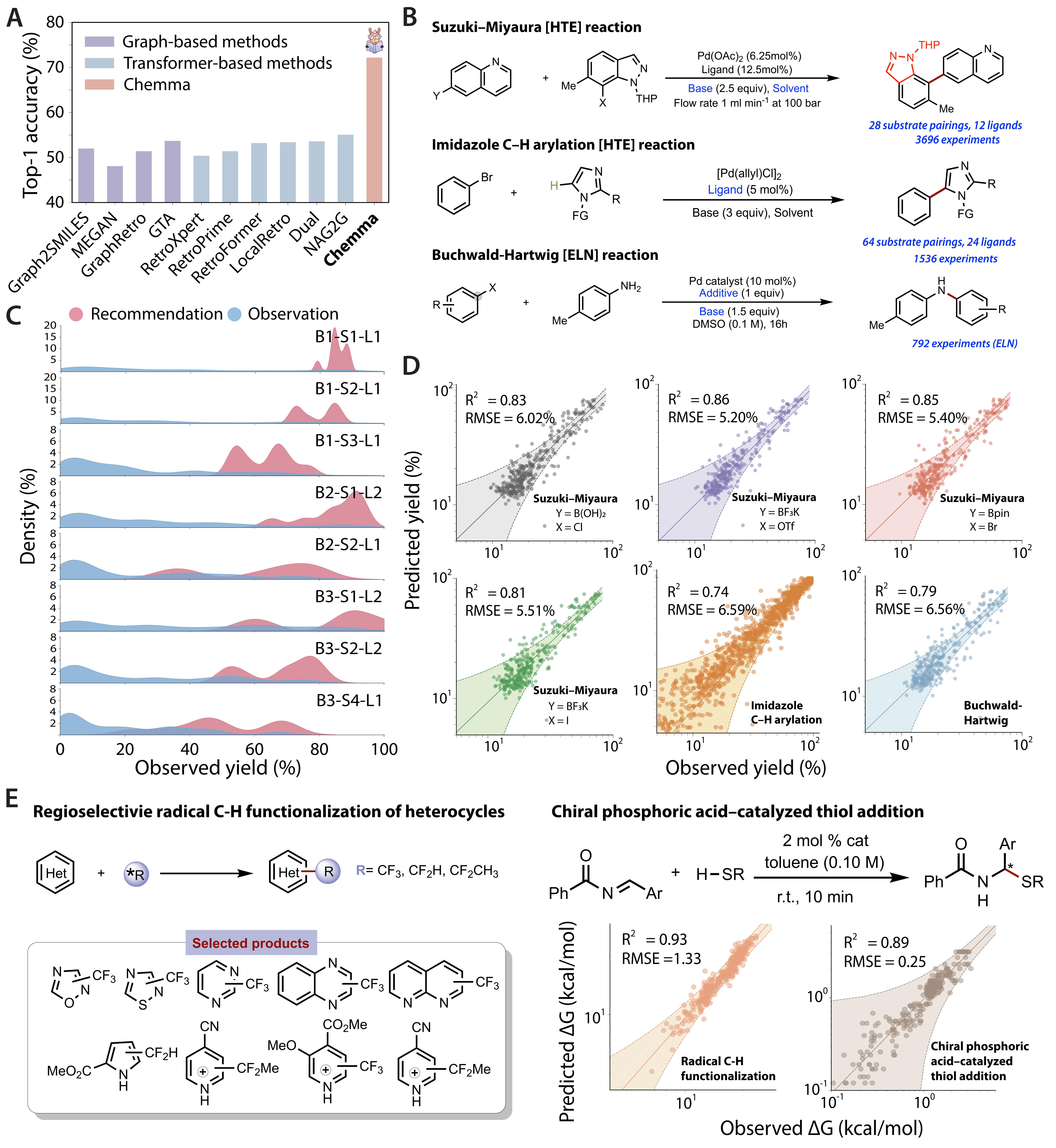}
\label{fig3: performance validation}
\end{figure*}
\noindent{{\bf Fig.~3. Performance evaluation of Chemma's capabilities for different organic synthesis tasks using both open benchmark and HTE datasets.} (\textbf{A}) Performance comparison of the one-step retrosynthesis task on the USPTO-50k test set. For a fair comparison, all of the competitive methods including Chemma are trained with the same USPTO-50k training set. The bars depict the top-1 accuracy of retrosynthesis and are color-coded by categories of the methods. 
(\textbf{B}) General schemes of the three Pd-catalysed reactions tested for condition recommendation and yield prediction. They are \suzuki~\cite{perera2018platform}, \bh~\cite{ruiz2016applications} and imidazole C--H arylation~\cite{wang2024identifying} reactions.
(\textbf{C}) Performance of ligand recommendation by Chemma on Pd-catalysed C--H arylation reaction. Given a reaction and its solvent and reagent, Chemma recommends a suitable ligand. Considering varying solvents and bases on \suzuki datasets, we depict yield distribution for each base-solvent-ligand (B-S-L) combination. The blue plot represents the distribution of observed yields of all ligands excluding the recommended one, while the red represents the yield distribution of the recommended ligand.
(\textbf{D}) Performance of yield prediction on Pd-catalysed \suzuki [HTE], \bh [ELN], and imidazole arylation [HTE] reactions. We conduct the validation in an out-of-sample fashion, randomly splitting a dataset into 70\% training and 30\% test sets. 
(\textbf{E}) Performance of selectivity prediction on two HTE datasets. We assess regioselectivity prediction on the regioselective radical C--H functionalization reactions, and enantioselectivity prediction on the chiral phosphoric acid-catalyzed thiol of addition \textit{N}-imides reactions. Scatter plots compare the predicted and observed $\Delta G$.}

\newpage
\begin{figure*}[htb!]
\centering
\includegraphics[width=1\linewidth]{Chemma_paper/figures/fig4-syntheticData.pdf}
\label{fig4: reaction optimization}
\end{figure*}
\noindent{{\bf Fig.~4. Value of Chemma-generated data for enhancing yield prediction and reaction optimization.} (\textbf{A}) Data organization for yield prediction. we randomly allocate a 10\% subset of datasets as the test set. We design seven scenarios for evaluation using varying fractions of real and synthetic data. For the other 90\% training data, we presume that only a fraction of data can be observed (e.g., scenario one, 5\%), and utilize Chemma to complete their yields (e.g., scenario one, 85\%). (\textbf{B}-\textbf{C}) Performance of yield prediction with varying fractions of real and generated data on Suzuki and Buchwald–Hartwig reaction, respectively. \textit{Chemma-enhanced RF} approach takes both real and generated data for training; \textit{RF without enhancement} approach only utilizes the real data. (\textbf{D}) Workflow of Chemma-BO for reaction optimization. Different from the classic Bayesian optimization, we use the Gaussian process surrogate model to fit the gap between the collected observation and the predicted yield by Chemma. (\textbf{E}-\textbf{F}) Averaged cumulative maximum observed yield with three methods, Chemma-BO, GPT-4, and BO, on Suzuki and Buchwald–Hartwig reaction, respectively. Initial experimental conditions are chosen randomly and five experiments are conducted per batch.}

%
\newpage
\begin{figure*}[htb!]
\centering
\includegraphics[width=1\linewidth]{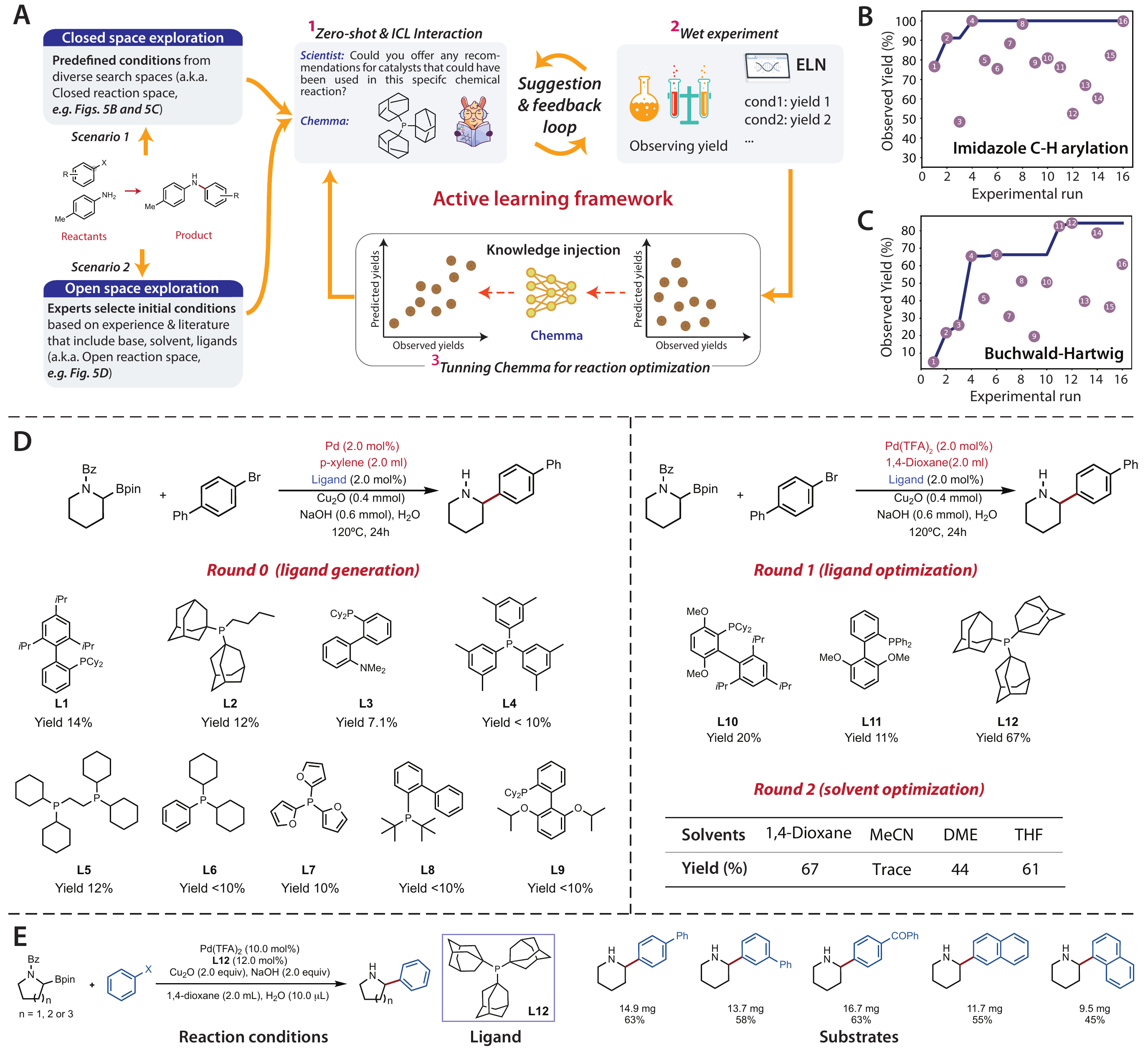}
\label{fig5: iterative optimization}
\end{figure*}
\noindent{{\bf Fig.~5. Illustrations of using Chemma to explore reaction spaces.}
(\textbf{A}) Overview of active learning framework for reaction exploration and optimization driven by Chemma. Given a new reaction, Chemma can work in two scenarios, (i) predefining reaction space, (ii) open reaction space. Next, in the ``suggestion \& feedback loop'', chemists follow Chemma's suggestion to conduct wet experiments. The results are stored in ELNs. After a round of experiments, if the target yield is not reached, chemists can fine-tune Chemma and start a new round of exploration.
(\textbf{B}-\textbf{C}) Optimization of ligand for the Pd-catalysed imidazole C--H arylation and \bh reactions on HTE data, respectively. Within the predefined reaction space, we first construct zero-shot prompts and ask Chemma to recommend a suitable ligand. We conduct 16 experimental runs for both two reactions. 
(\textbf{D}) Exploring open reaction space with Chemma for an unreported reaction, synthesis of Pd-catalysed N-heterocycles. Nine ligands selected by experts' experience and Chemma's suggestion constitute the initial condition scope, defined as \textit{round 0}. After that, we update Chemma by RLHF. We then ask Chemma to generate three possible ligands by interacting with chemists, and conduct wet experiments in \textit{round 1}. Next in \textit{round 2}, we test three more solvents suggested by Chemma via ICL prompts. (\textbf{E}) Test for substrates' compatibility. PAd\textsubscript{3}/Pd-catalyzed C--C coupling of (1-benzoylpiperidin-2-yl)boronic acid ester and diverse aryl bromides}. The selected reaction conditions are 1,4-dioxane (2 mL), Cu\textsubscript{2}O, NaOH (0.6 mmol) performed with a \SI{120}{\degreeCelsius}.


\clearpage

\begin{figure}[!h]
\label{fig:S1}
   \begin{center}
  \includegraphics[width=0.8\textwidth]{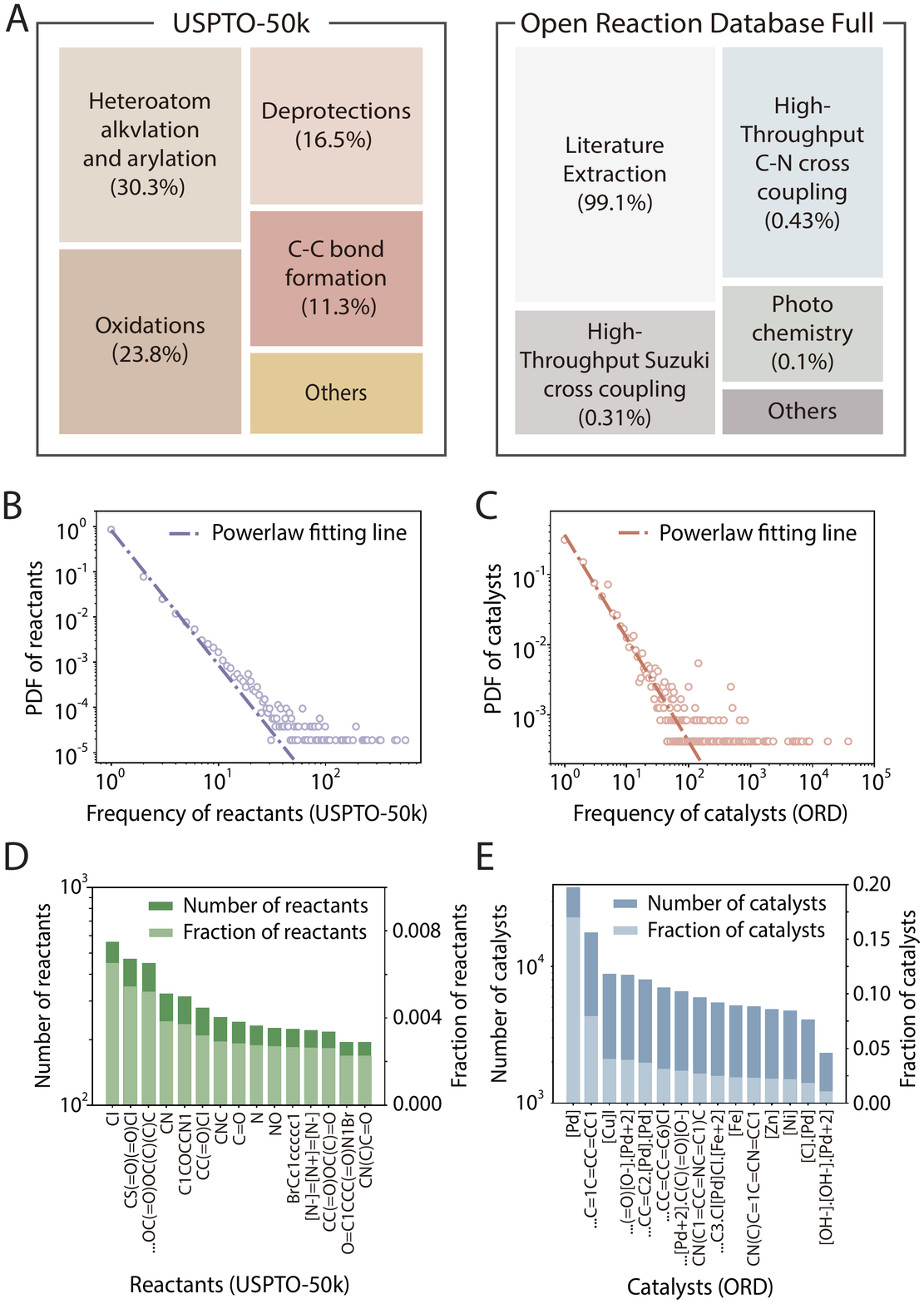}
   \end{center}
   \vspace{-3mm}
\end{figure}
\noindent \textbf{Extended Data Fig.~1. Distribution of types of reactions in the USPTO-50k and ORD datasets.} (\textbf{A}) Data organization of USPTO-50k and ORD datasets. All of reactions in USPTO-50k are from patents in the United States; Most of reactions in ORD are from literature. (\textbf{B}-\textbf{C}) Power law fitting of the reactant distribution in the USPTO-50k and the catalyst distribution in the ORD, where the shallow points show the probability density and the deep dashed-line shows the ideal power-law fitting, respectively. (\textbf{D}-\textbf{E}) The bar charts of fifteen most common reactants and catalysts in the USPTO-50k and ORD, respectively. The shallow color presents the decimal-scale proportion and the deep color presents the log-scale count.

\clearpage
\begin{figure*}[!h]
   \begin{center}
  \includegraphics[width=0.85\textwidth]{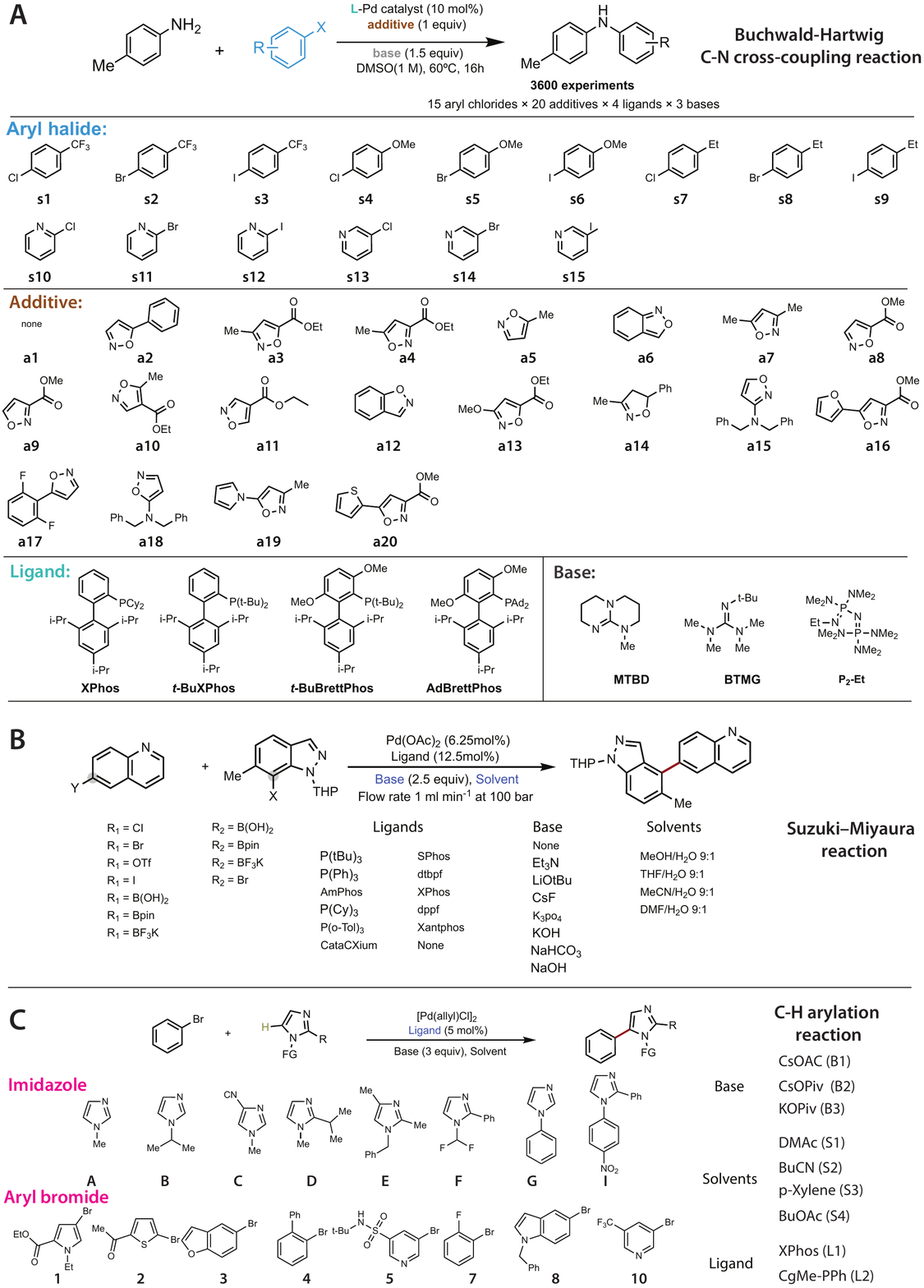}
   \end{center}
   \vspace{-3mm}
  \label{fig:S2}
\end{figure*} 
\noindent \textbf{Extended Data Fig.~2. Description of the three HTE datasets.} (\textbf{A}) Pd-catalysed Buchwald-Hartwig C--N coupling reaction: aryl halides, isoxazole additives, Pd precatalyst, ligands and bases. (\textbf{B}) \suzuki reaction, consisting of the reaction yield measured as a function of boronic acid derivative, aryl halide, ligand, base and solvent. (\textbf{C}) C–H arylation dataset components: ligands, imidazoles, aryl bromides.

\clearpage
\begin{figure*}[!h]
   \begin{center}
  \includegraphics[width=1.0\textwidth]{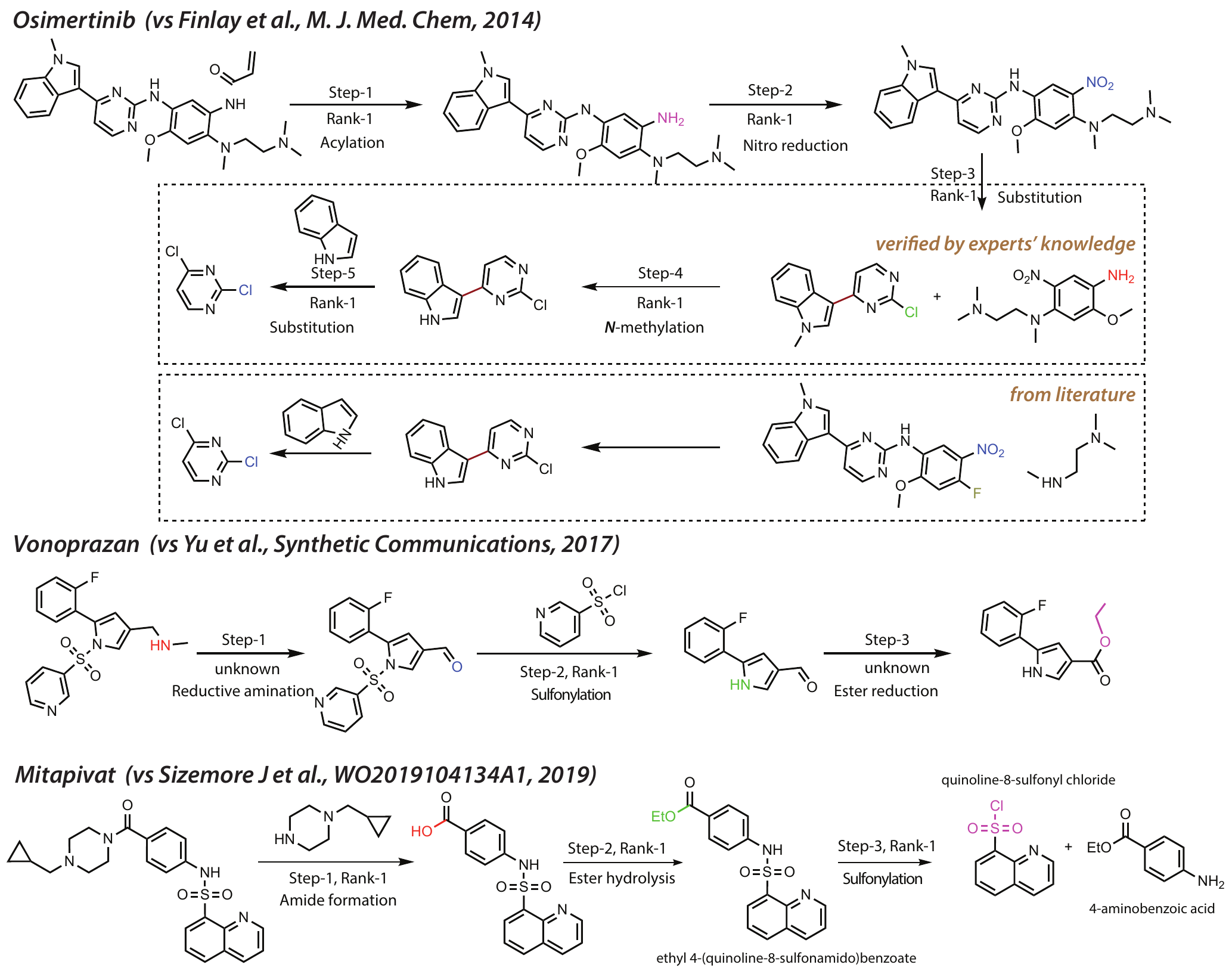}
   \end{center}
   \vspace{-3mm}
  \label{fig:case-study-retro}
\end{figure*}

\clearpage
\begin{figure*}[!h]
   \begin{center}
  \includegraphics[width=1.0\textwidth]{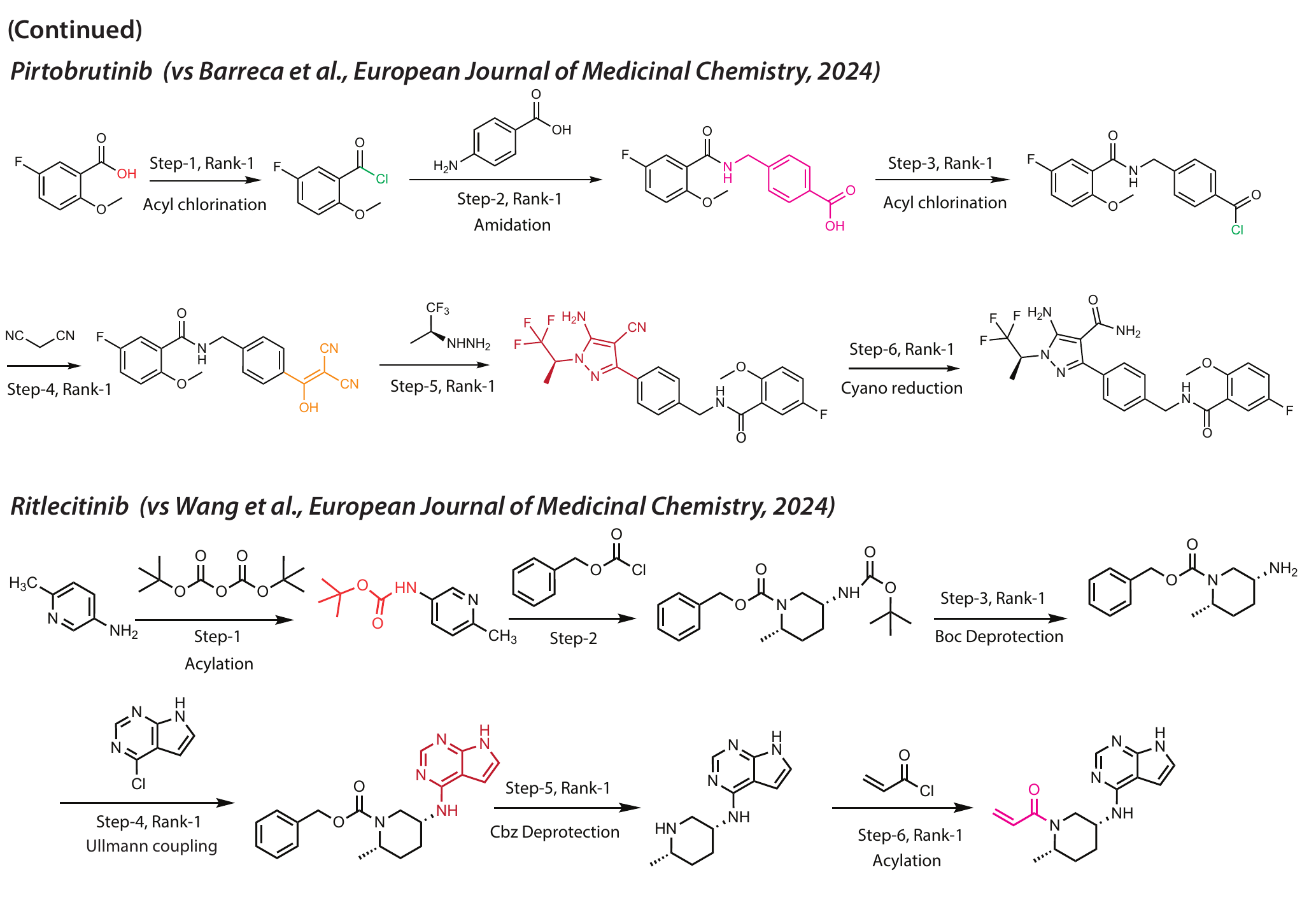}
   \end{center}
   \vspace{-3mm}
  \label{fig:case-study-retro-2}
\end{figure*}
\noindent \textbf{Extended Data Fig.~3. Multi-step retrosynthesis routes of five drug molecules predicted by Chemma.} We select five drug molecules as the target products including Osimertinib, Vonoprazan, Mitapivat, Pirtobrutinib, and Ritlecitinib. The reaction centers and leaving groups are highlighted in different colors. ``Rank-1'' in predicted synthetic routes indicates that our predicted step with the highest probability hits the synthetic route reported in the literature.

\clearpage
\begin{figure*}[!h]
   \begin{center}
  \includegraphics[width=1\textwidth]{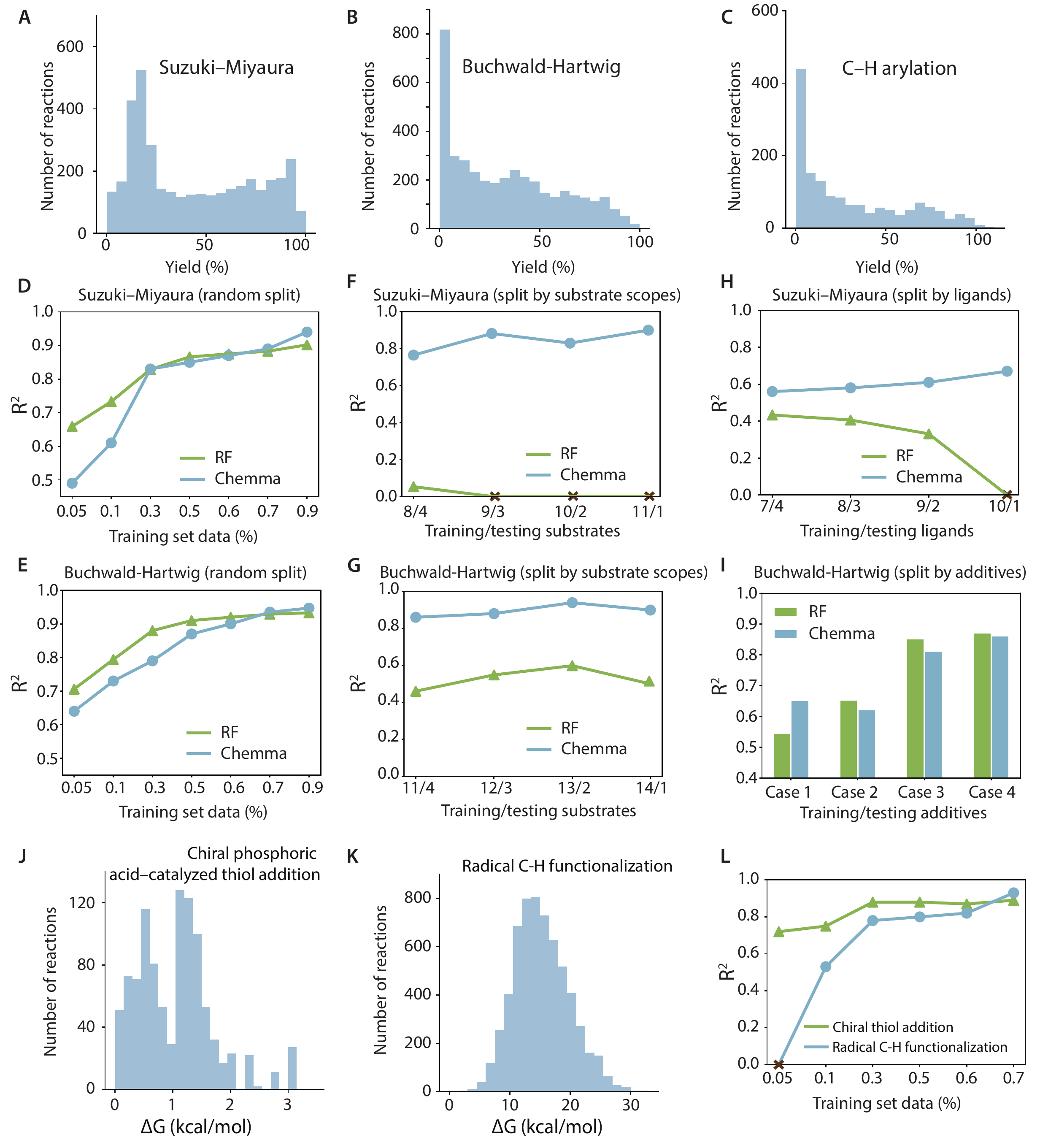}
   \end{center}
   \vspace{-3mm}
  \label{fig:HTE-dataset-distribution}
\end{figure*}
\noindent \textbf{Extended Data Fig.~4. Assessment of yield prediction performance by RF model and Chemma on two HTE reactions: Suzuki-Miyaura and Buchwald-Hartwig.} (\textbf{A-C}) The distribution of the yields for Pd-catalysed Suzuki–Miyaura [HTE], Buchwald-Hartwig [ELN], and imidazole C--H arylation [HTE] reactions. (\textbf{D, E}) Test set performance of the RF model and Chemma with randomly split strategy. A gradual erosion in predictive accuracy occurred from 90\% of the entire dataset down to 5\% of the full data set. Each training set with different proportions are selected randomly from the original full data set. \textbf{(F, G)} Test set performance of the RF model and Chemma when training and test sets are split by diverse substrate scopes. For Suzuki-Miyaura reaction, all reactions can be split by twenty different kinds of substrates. For the Buchwald-Hartwig reaction, all reactions are split by fifteen aryl chloride substrates. We define four scenarios for evaluation characterized by variable training and testing substrates. For instance, a scenario encompasses reactions involving eight substrates in the training phase and reactions with four substrates for testing. Ten-cross validation is used for fair comparison. (\textbf{H, I}) Test set performance of the RF model and Chemma by isolating conditions sets. For the Suzuki-Miyaura reaction, all reactions are split by eleven different kinds of ligands. For the Buchwald-Hartwig reaction, we select four case scenarios that had been tested in~\cite{ahneman2018predicting} for evaluation. Training and testing sets are divided with additives scopes. The indices of tested additives are marked in Extended Data Fig.~2A. Accordingly, in Case-1, tested additives are enumerated by indices a10, a18, a15, a23, and a4; in Case-2, indices a11, a9, a1, a17 and a5 are selected for testing; in Case-3, indices a14, a8, a21, a12, a6 are selected for testing; in Case-4, indices a16, a2, a22, a20, a3 are selected for testing. (\textbf{J-K}) The distribution of free energy barriers on two reactions: chiral phosphoric acid-catalyzed thiol addition and radical C--H functionalization reaction. (\textbf{L}) Test set performance by Chemma on two selective datasets. Each training sets with different proportions are selected randomly from the original full data set, and 30\% data of the entire datasets are randomly selected as test sets.

\clearpage
\begin{figure*}[!h]
   \begin{center}
  \includegraphics[width=1\textwidth]{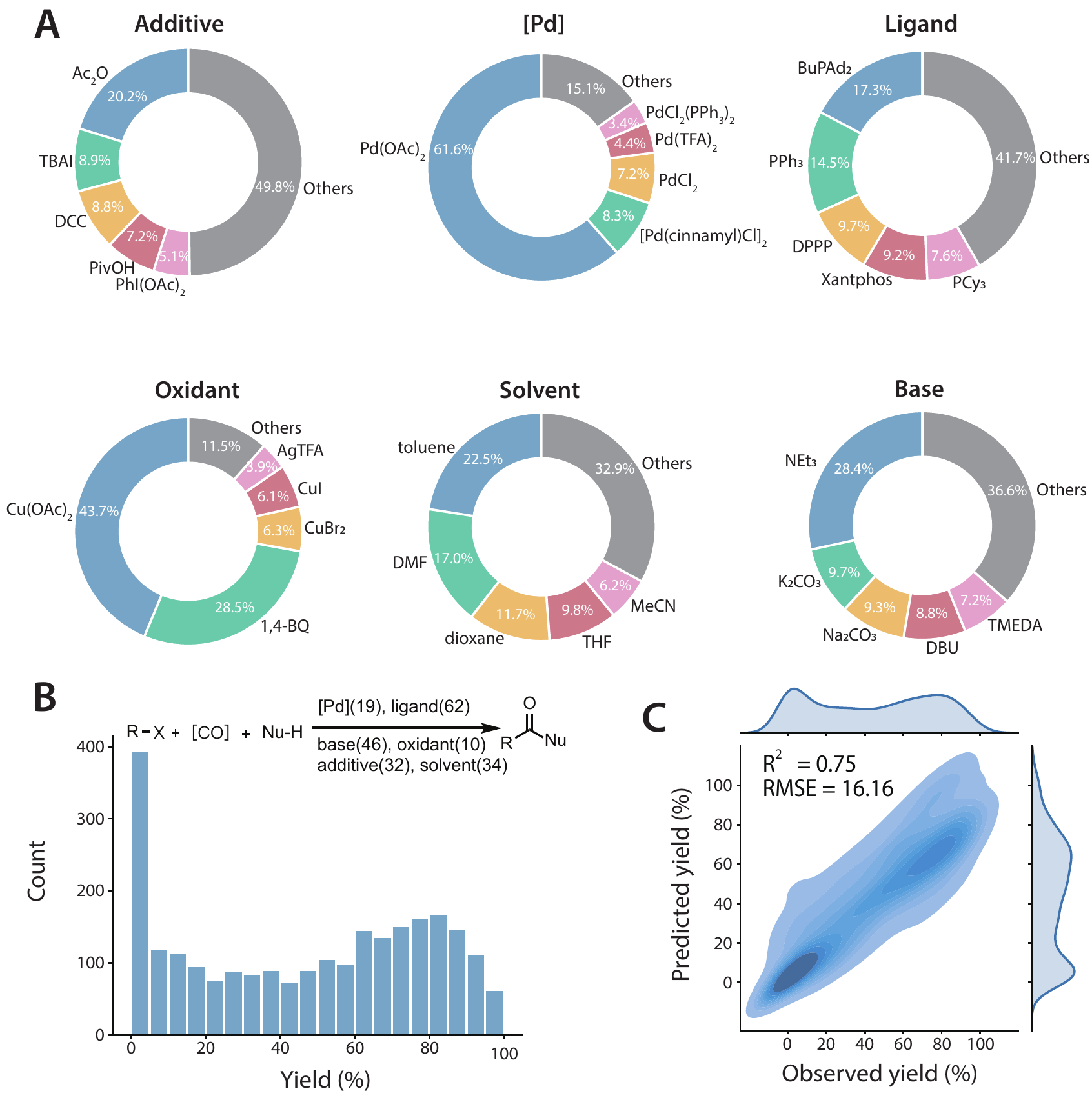}
   \end{center}
   \vspace{-3mm}
  \label{fig:pd-cata-carbon}
\end{figure*}

\noindent \textbf{Extended Data Fig.~5. Performance evaluation of Chemma for yield prediction with literature-derived Pd-catalyzed homogeneous carbonylation reactions}. (\textbf{A}) Proportions of the top five most frequently used catalyst precursors, ligands, bases, oxidants, additives, and solvents in the data set. \textbf{(B)} Schematic representation and yield distribution of the Pd-catalyzed carbonylation reaction.  \textbf{(C)} Predictive results of the Chemma. We assess the generalization capability of Chemma in the out-of-sample testing strategy. Within the training set, we select literature IDs $\leq$ 100, while the testing set exclusively included data from the original testing set with literature IDs \textgreater 100, challenging the Chemma to predict reaction performance in chemical spaces it has not previously encountered.

\clearpage
\begin{figure*}[!h]
   \begin{center}
  \includegraphics[width=1.0\textwidth]{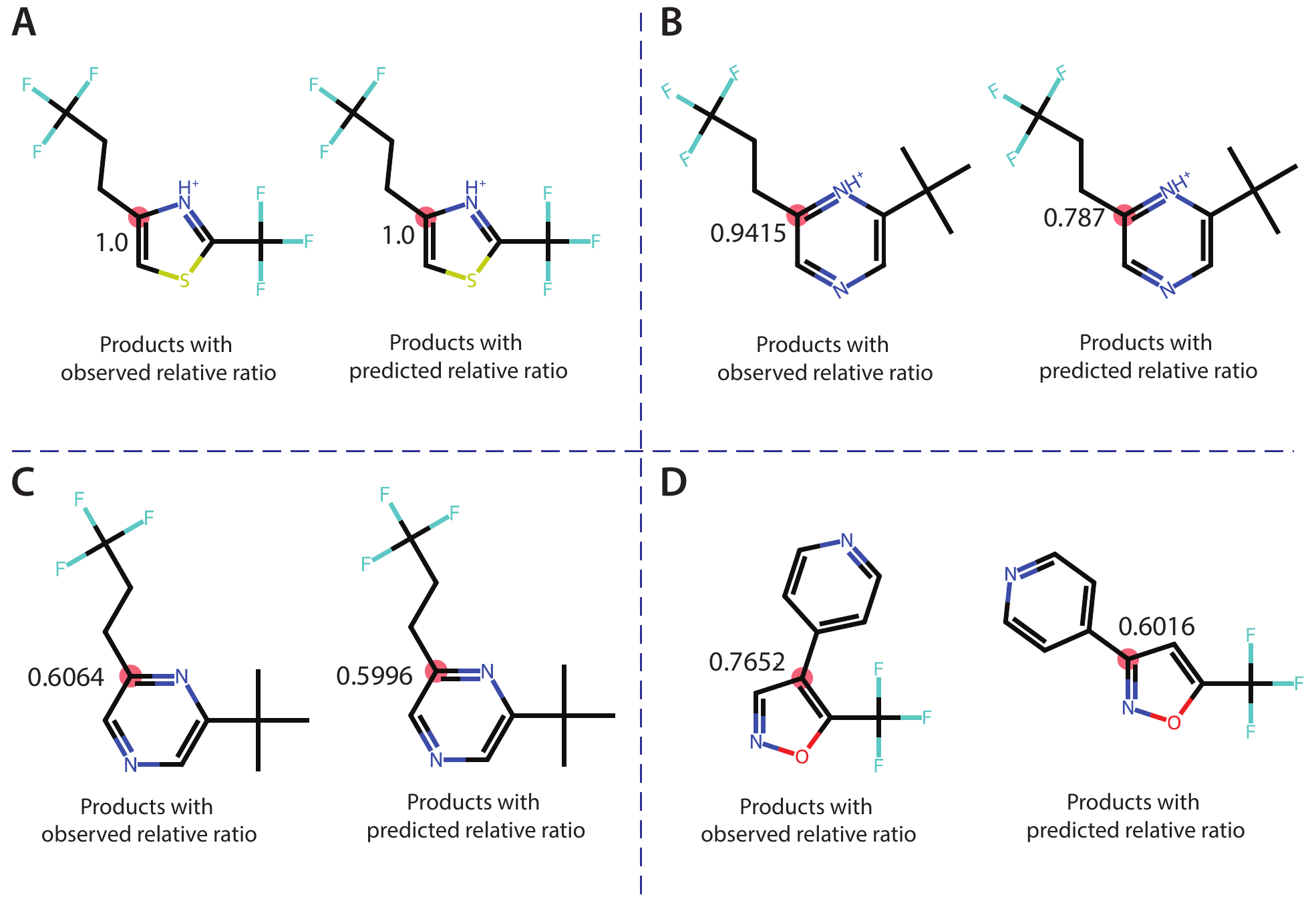}
   \end{center}
   \vspace{-3mm}
  \label{fig:case-study-selectivity}
\end{figure*}
\noindent \textbf{Extended Data Fig.~6. Case study and visualization of selectivity prediction performance.} (\textbf{A-C}) Successful case: comparison of sites predictions and experimentally determined for products. For each reaction, the predicted site is exactly same with the labeled one. Reactive sites of products are highlighted with a circle. (\textbf{D}) Illustration of failure prediction results. The measured probability of the target product with N m-meta site is 0.7652, but the predicted result is 0.6016 with N o-ortho site.

\clearpage

\begin{figure*}[!h]
   \begin{center}
  \includegraphics[width=1.0\textwidth]{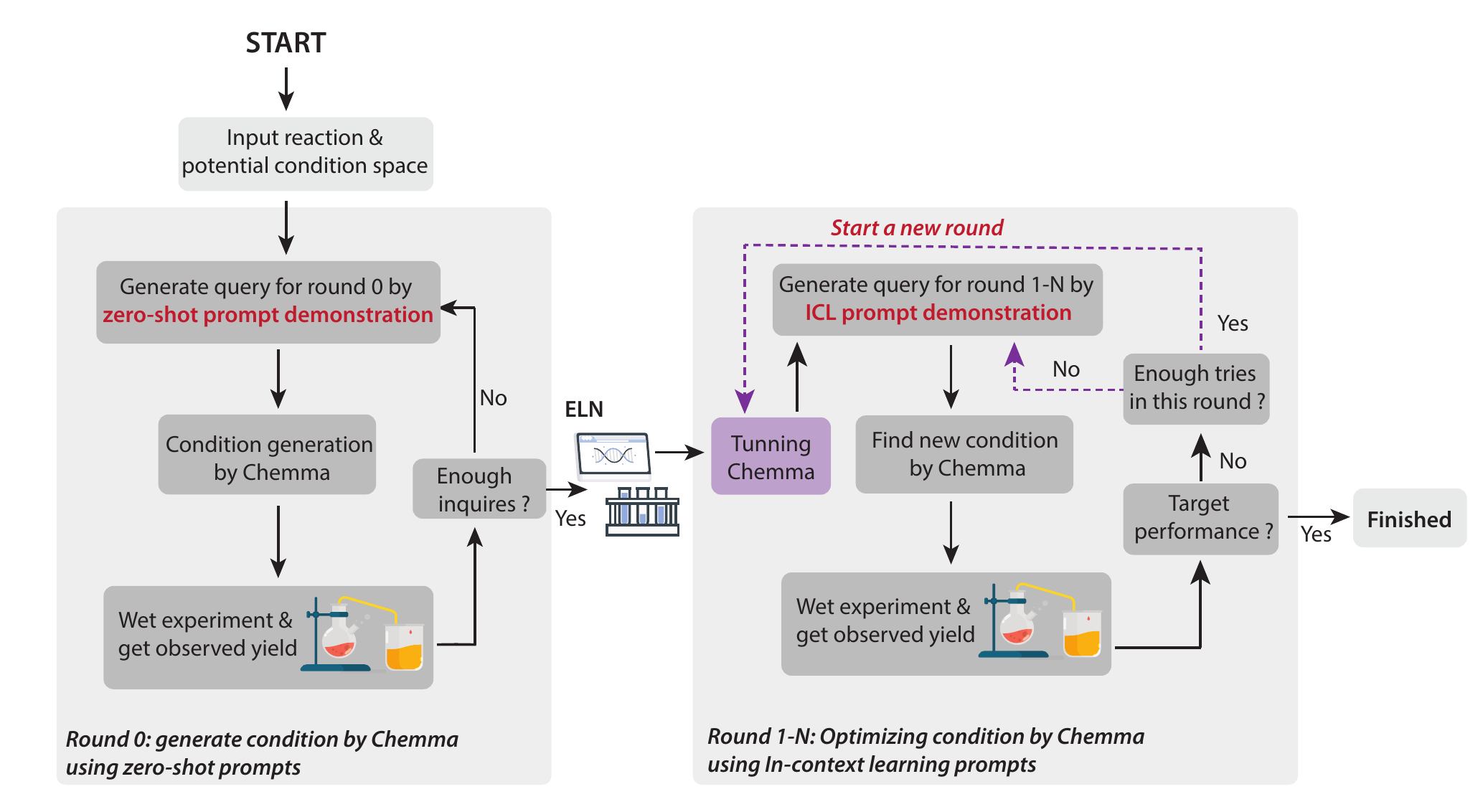}
   \end{center}
   \vspace{-3mm}
  \label{fig:workflow of active learning}
\end{figure*}
\noindent \textbf{Extended Data Fig.~7. Detailed workflow of active learning framework for reaction exploration and optimization driven by Chemma.} We position Chemma as a chemistry assistant and integrate it into an active learning framework for reaction exploration and optimization. In \textit{round 0}, Chemma iteratively suggests the next reaction condition considering the feedback of last wet experiment. After a round of ``suggestion-feedback loop'', we can enter into \textit{rounds 1-N}, and fine-tune Chemma to adapt the reactions. The framework works not only on pre-defined reaction spaces by experts, a.k.a., closed reaction space, but also on open reaction space where the conditions are not limited to experts' prior knowledge.

\clearpage
\begin{figure*}[!h]
   \begin{center}
  \includegraphics[width=1.0\textwidth]{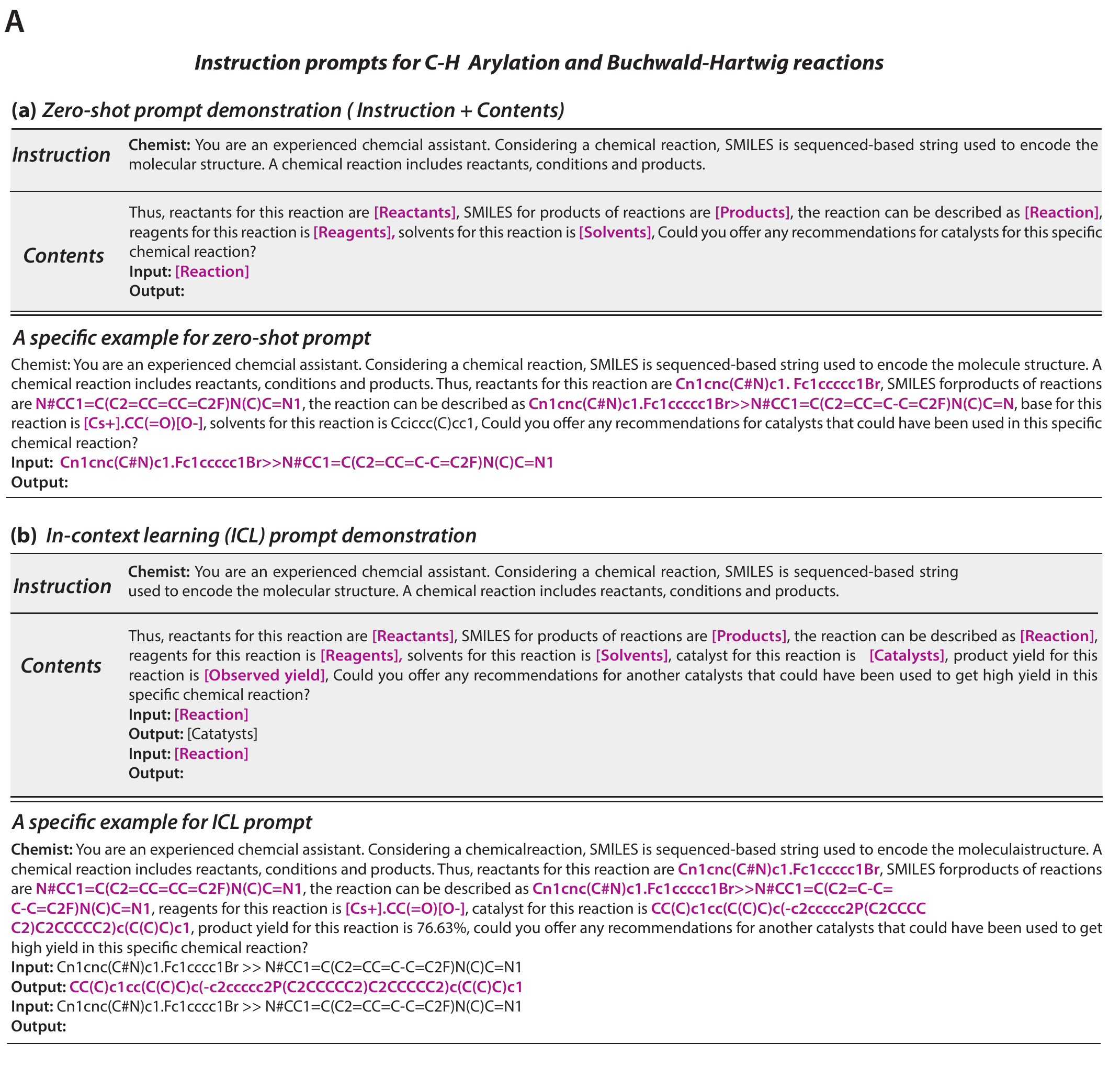}
   \end{center}
   \vspace{-3mm}
  \label{fig:S5}
\end{figure*}

\clearpage
\begin{figure*}[!h]
   \begin{center}
  \includegraphics[width=1.0\textwidth]{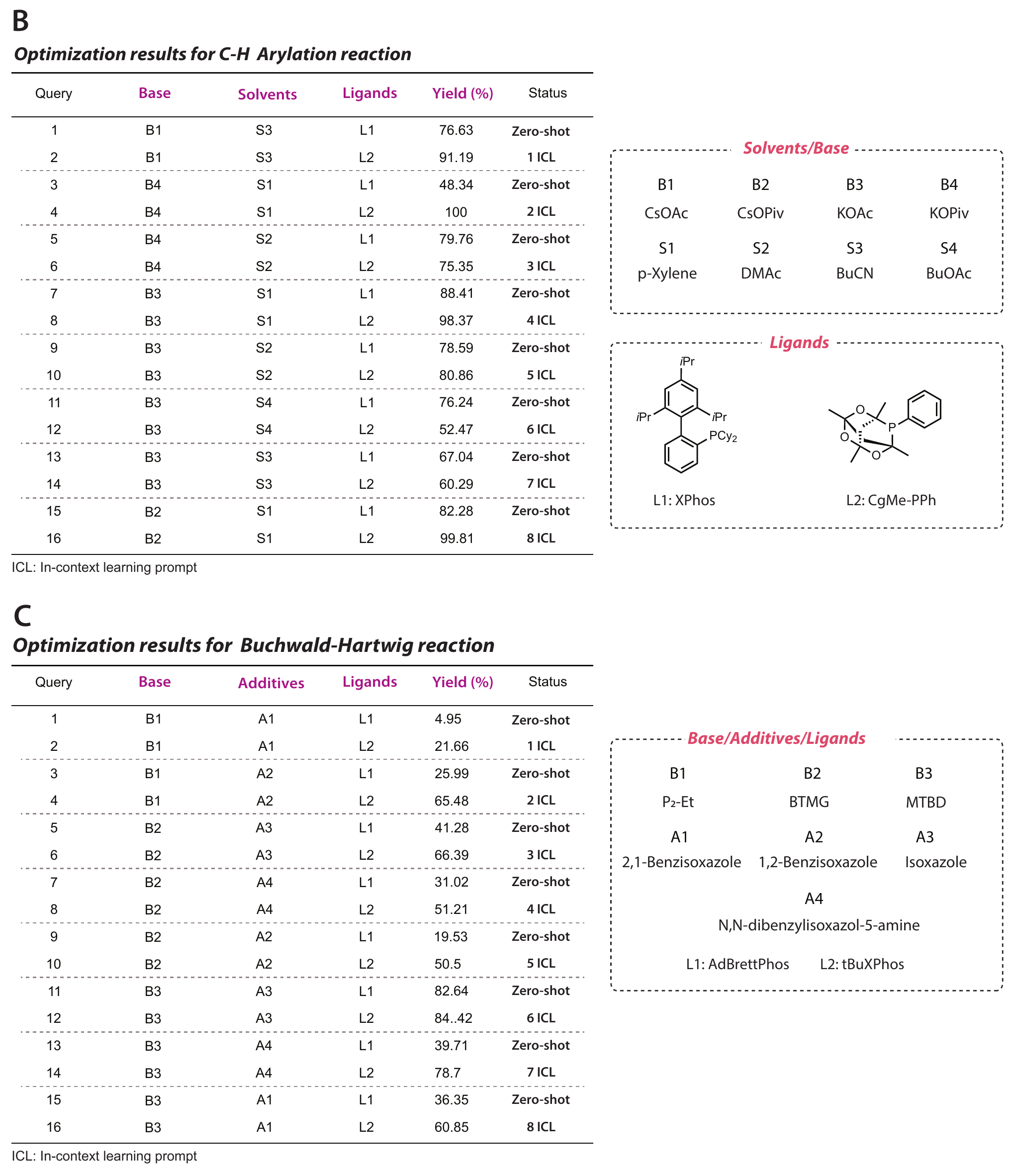}
   \end{center}
   \vspace{-3mm}
  \label{fig:si-fig5b-table}
\end{figure*}

\noindent \textbf{Extended Data Fig.~8. Illustration of detailed prompts designed as the input of Chemma for optimizing reactions including Pd-catalysed imidazole C--H arylation reaction, Pd-catalysed Buchwald-Hartwig reaction, and our unreported 
synthesis of aryl-substituted reaction of nitrogen heterocycles, respectively.} (\textbf{A}) The detailed prompts used for optimizing conditions of imidazole C--H arylation and Buchwald-Hartwig reactions. (\textbf{B-C}) Details of reaction optimization process. A total of 16 experiments are conducted in s closed-loop fashion continuously. Taking imidazole C--H arylation reaction as a example, for each optimized process, the initial solvent-base variables are randomly selected from the entire reaction space. We interact with Chemma by zero-shot prompts to acquire a suitable ligand. Initial generation provides a preliminary exploration of the reaction space. Subsequently, we leverage both the observed yield from the initial exploration and the generated ligand to construct the ICL prompts and ask Chemma to suggest a ``higher-yield'' ligand. If the suggested ligand has been tested in previous runs, we randomly change the reaction condition variables except the ligand for the next group of zero-shot interaction and experimental runs.

\clearpage
\begin{figure*}[!h]
   \begin{center}
  \includegraphics[width=1.0\textwidth]{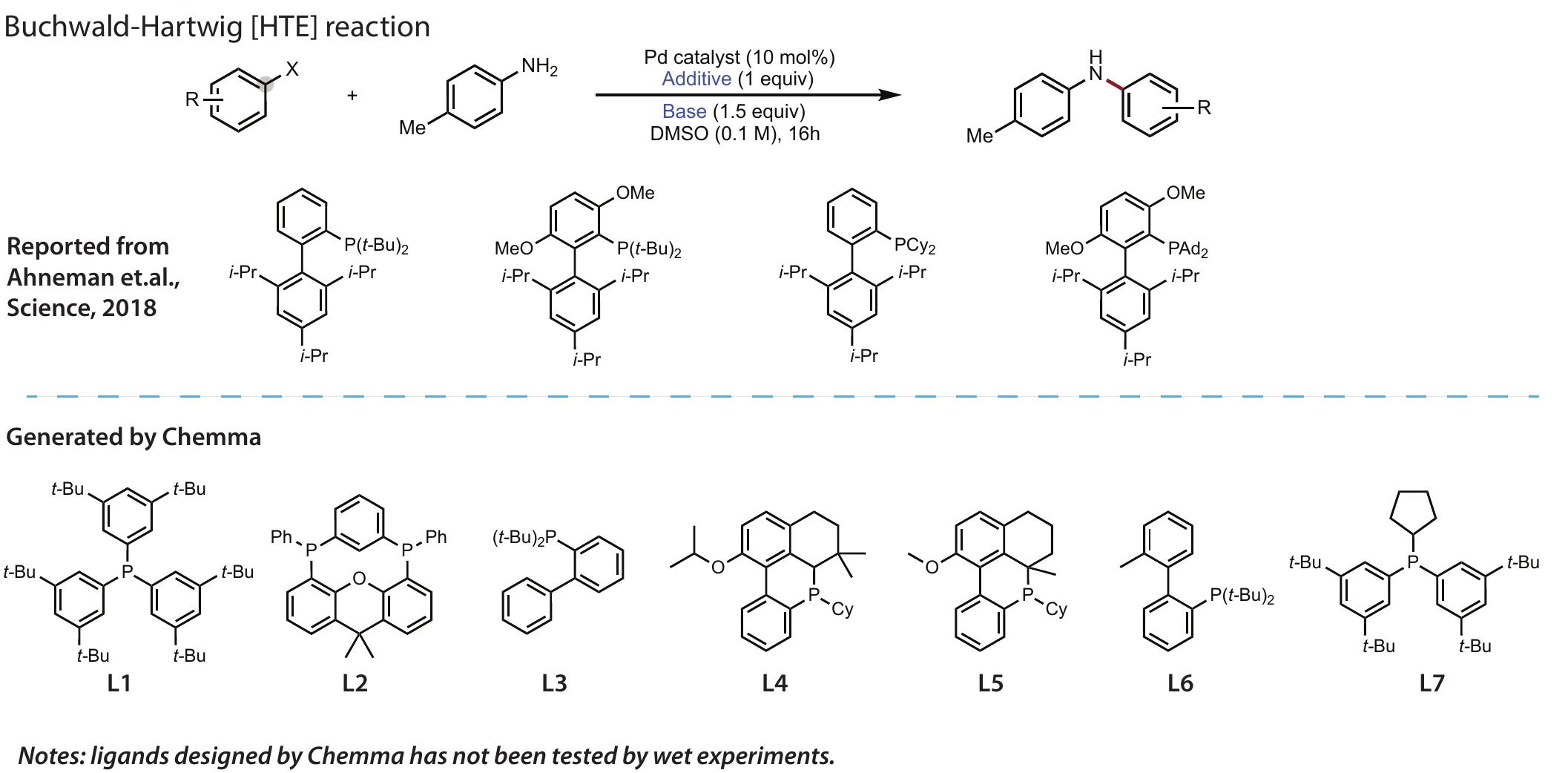}
   \end{center}
   \vspace{-3mm}
  \label{fig:S8}
\end{figure*}
 \noindent \textbf{Extended Data Fig.~9. The new ligands designed by Chemma for the Pd-catalysed \bh reaction.} For a reported HTE reaction, we ask Chemma to design new ligands (L1 to L7) that have not been explored. It is worth noting that we synthesize L1, L3, and L6, and conduct wet experiments to evaluate the effectiveness of the generated ligands. The yield of L1 and L3 is 6\% and 16\%, respectively. L6 exhibits no reactivity.

\clearpage
\begin{figure*}[!h]
   \begin{center}
  \includegraphics[width=1.0\textwidth]{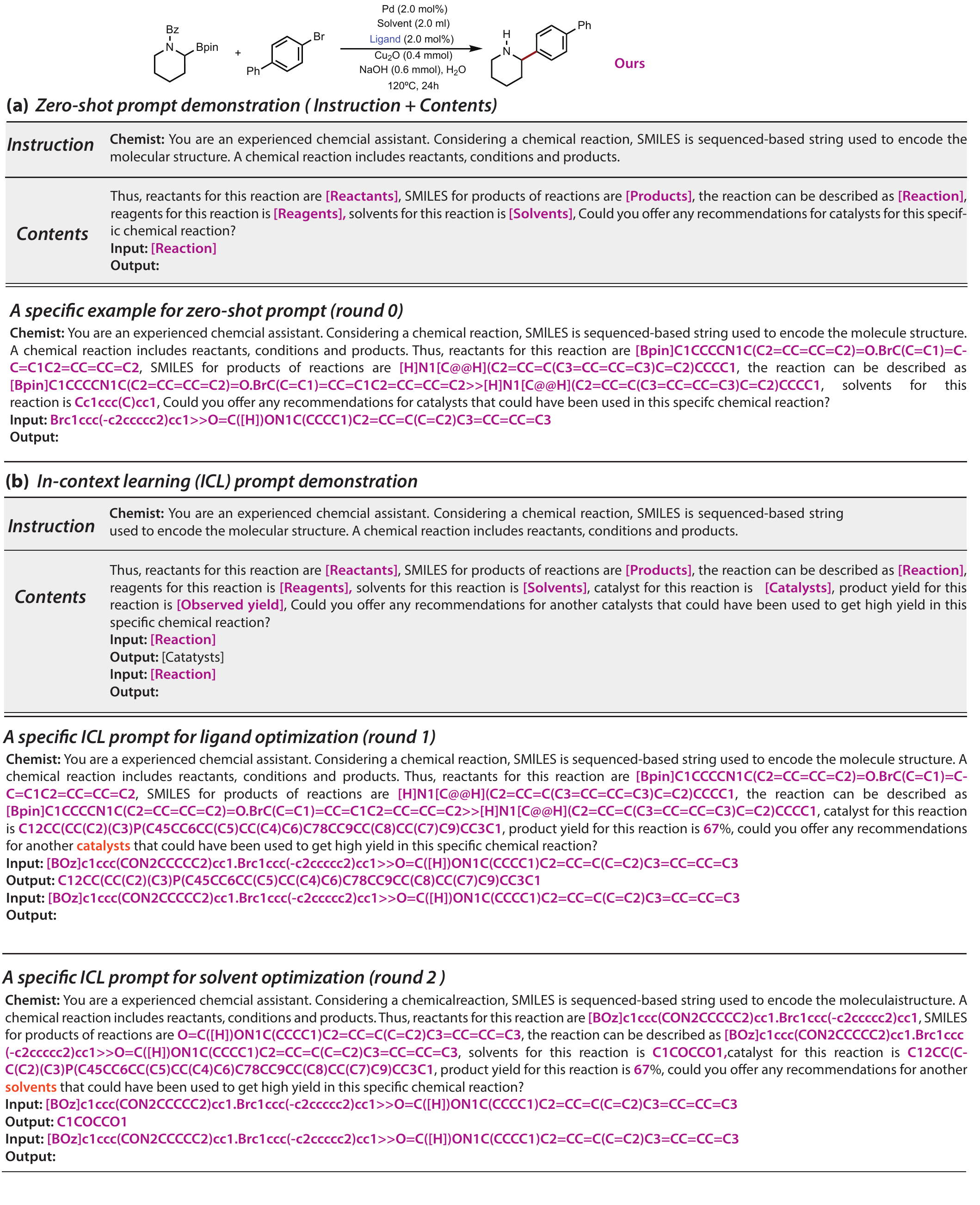}
   \end{center}
   \vspace{-3mm}
  \label{fig:SI-Fig5D}
\end{figure*}
 \noindent \textbf{Extended Data Fig.~10.  Illustration of detailed prompts designed as input to the Chemma for optimizing the reported synthesis of aryl-substituted reaction of nitrogen heterocycles.}

\end{document}